# Light-Driven Nanoscale Vectorial Currents

Jacob Pettine[1]*, Prashant Padmanabhan[1], Teng Shi[1], Lauren Gingras[2], Luke McClintock[1,3], Chun-Chieh Chang[1], Kevin W. C. Kwock[1,4], Long Yuan[1], Yue Huang[1], John Nogan[5], Jon K. Baldwin[1], Peter Adel[2], Ronald Holzwarth[2], Abul K. Azad[1], Filip Ronning[6], Antoinette J. Taylor[1], Rohit P. Prasankumar[1,7], Shi-Zeng Lin[1], and Hou-Tong Chen[1]*

[1] Center for Integrated Nanotechnologies, Los Alamos National Laboratory, Los Alamos, NM 87545, United States

[2] Menlo Systems, Martinsried, Bavaria, 82152 Germany

[3] Department of Physics, University of California-Davis, Davis, CA 95616, United States

[4] The Fu Foundation School of Engineering and Applied Science, Columbia University, New York, NY 10027, United States

[5] Center for Integrated Nanotechnologies, Sandia National Laboratories, Albuquerque, NM 87123, United States

[6] Institute for Material Science, Los Alamos National Laboratory, Los Alamos, NM 87545, United States

[7] Intellectual Ventures, Bellevue, WA 98005, United States

* jacob.pettine@lanl.gov, chenht@lanl.gov

**Controlled charge flows are fundamental to many areas of science and technology, serving as carriers of energy and information, as probes of material properties and dynamics[1], and as a means of revealing[2,3] or even inducing[4,5] broken symmetries. Emerging methods for light-based current control[5-15] offer promising routes beyond the speed and adaptability limitations of conventional voltage-driven systems. However, optical generation and manipulation of currents at nanometer spatial scales remains a basic challenge and a crucial step towards scalable optoelectronic systems for microelectronics and information science. Here, we introduce vectorial optoelectronic metasurfaces in which ultrafast light pulses induce local directional charge flows around symmetry-broken plasmonic nanostructures, with tunable responses and arbitrary patterning down to sub-diffractive nanometer scales. Local symmetries and vectorial current distributions are revealed by polarization- and wavelength-sensitive electrical readout and terahertz (THz) emission, while spatially-tailored global currents are demonstrated in the direct generation of elusive broadband THz vector beams[16]. We show that in graphene, a detailed interplay between electrodynamic, thermodynamic, and hydrodynamic degrees of freedom gives rise to rapidly-evolving nanoscale driving forces and charge flows under extreme temporal and spatial confinement. These results set the stage for versatile patterning and optical control over nanoscale currents in materials diagnostics, THz spectroscopies, nano-magnetism, and ultrafast information processing.**

Recent advances in controlling photocurrents (currents induced by light fields) have been responsible for a variety of new insights in areas ranging from material characterization and device physics[1] to electrolytic chemistry[17] and ultrafast electron diffraction and imaging[18]. The evolution from voltage-driven to light-driven processes has been particularly prominent in information science and microelectronics, where optoelectronic and opto-spintronic currents in emerging topological[5,10,11], magnetic[12,13], and low-dimensional[14,19] materials are introducing faster speed limits and light-based control degrees of freedom. However, the broken spatial or temporal symmetries responsible for photocurrent generation in these materials[3] are typically either intrinsic to the lattice and thus constrained to specific light–matter interaction geometries, or otherwise dependent upon applied static fields that are difficult to texture on small length scales. While coherent lightwave interactions in strong, few-cycle, and/or phase-stabilized light fields[6-9] transcend many material-specific challenges and offer a high degree of control over charge motion, such approaches have also remained limited to single laser spots or larger micrometer-scale structured light fields.

Plasmonic systems can be utilized to overcome these spatial limitations by concentrating light down into deeply sub-diffractive nanometer scales. Such systems are already known to provide exceptional control over energy flow between photonic, electronic, and thermal degrees of freedom on ultrafast timescales[20]. A better understanding of momentum flow in plasmon-excited hot carrier distributions is also emerging[21], as determined by spatially-tailored plasmonic hot spots and directionality imposed by nanoscale geometry. Although such effects have primarily been explored via nonlinear photoemission into free space[22-24], linear photocurrent responses under continuous-wave excitation have been observed recently in hybrid plasmonic systems serving as bias-free mid-infrared photodetectors[25,26] as well as spin-valley polarized valleytronic transistors[27].

Here, we show that plasmonic metasurfaces offer far broader capabilities for harnessing charge flow at nanometer spatial scales and femtosecond time scales. Photocurrents are a universal manifestation of broken inversion symmetry[3], and we find that asymmetric gold nanoantennas on graphene exhibit strong light-driven directional responses. Local directionality is determined by the orientation of individual nanoantennas, with general implications for global spatially-varying and optically-controlled photocurrents. As an immediate application, we demonstrate that these vectorial optoelectronic metasurfaces serve as efficient and versatile sources of ultrafast THz radiation, including broadband THz vector beams. Electrostatic gating and multiphysics modeling reveal a local photothermoelectric driving mechanism and elucidate previously unexplored dynamics occurring at the intersection of femtosecond excitation and nanoscale localization.

**Generation of Ultrafast Directional Photocurrents**

Our optoelectronic metasurface concept is illustrated in Fig. 1a, where inversion symmetry-broken gold nanoantennas act as lightning rods[28] with strong resonantly-enhanced light fields at their sharp tips (~15 nm radius; Fig. 1b,c insets). Hot carrier excitation at these nanoplasmonic hot spots drives local current density $\mathbf{j}(\mathbf{r},t)$ within the underlying graphene. The broken inversion symmetry leads to net current along the orientation of the nanoantennas, which decays on sub-picosecond timescales due to momentum-relaxing scattering with optical phonons, impurities, and substrate phonons[29-31]. These transient photocurrents radiate free-space THz waves, $\mathbf{E}_{\text{THz}} \propto -d\mathbf{j}/dt$, which are utilized as a contact-free probe of the ultrafast current dynamics. Additionally, the time-averaged DC photocurrent is read out directly via electrical contacts for unambiguous verification of the overall charge flow behaviors.

We first investigate the photocurrent response in metasurfaces with uniform nanoantenna orientation and thus a globally preferred directionality. The plasmonic resonance is broadly tunable from visible to infrared wavelengths based on the nanoantenna structure (shape and size), material, and dielectric environment. Structural tuning is exploited here for resonances at 800 nm and 1550 nm, as clearly observed in the measured and simulated transmission spectra (Fig. 1b). Corresponding enhancements of the ultrafast THz fields and DC photocurrents are shown in Fig. 1c, in good agreement with the simulated field intensity enhancements, which underscores the central role of the (asymmetric) plasmonic fields in the photocurrent generation process.

Single-cycle THz pulses are observed from both the 800 nm and 1550 nm (Fig. 1d) metasurfaces via free-space electro-optic sampling (Methods), indicating the photocurrent rise and decay on few-hundred femtosecond timescales. These pulses disappear when the exposed graphene is etched away without removing the nanoantennas (Supplementary Note 1), verifying that the THz radiation originates from graphene photocurrents rather than optical rectification directly at the nanoantennas[32]. Remarkably, with little optimization (Supplementary Note 2), the 30-nm-thick metasurfaces yield THz fields that are comparable to those from widely-utilized 1-mm-thick ⟨110⟩ ZnTe nonlinear crystals under the same excitation conditions. This indicates the potential utility of these optoelectronic metasurfaces as new THz sources with versatile responses arising from sample orientation and patterning as well as incident polarization and wavelength, as explored further below.

A linear dependence on incident fluence (Fig. 1e) is observed in both the THz field amplitude and photocurrent readout for the 800 nm metasurface below ~0.8 µJ cm$^{-2}$, which can result from various possible mechanisms at metal-graphene junctions, such as photovoltaic[14,33] or photo-Dember[34] effects. However, photothermoelectric effects are generally found to prevail, including under continuous-wave excitation at the nanoscale[25,26] and ultrafast excitation at mesoscopic metal-graphene junctions[35,36]. We will later show that photothermoelectric effects remain dominant under simultaneous ultrafast, nano-localized excitation, but we first explore the general implications of artificially patterned linear photocurrent responses, which can be extended to other hybrid material systems and are independent of the specific mechanism.

**Local Vectorial Photocurrents and Global Responses**

For metasurfaces with uniform spatial patterning as shown in Fig. 2a,b, global current and THz emission measurements reveal the local light-matter interaction symmetries and photocurrent responses at the nanoscale unit cell level. The oriented metasurface (*pm* wallpaper group, Fig. 2a) exhibits a particularly simple $\cos^2(\theta)$ dependence on the incident linear polarization angle with respect to the nanoantenna principal axis of broken symmetry (Fig. 2c), which is characteristic of the linear dependence on the projected field intensity. The photocurrent directionality remains along this axis, regardless of the incident light polarization. Under circularly-polarized incident light, the overall nanoantenna-oriented photocurrent across the metasurface is revealed by the spatially-uniform linear polarization of the radiated THz beam, which is continuously rotated by sample rotation (Extended Data Fig. 1).

Additional symmetry can be introduced into the metasurface unit cell with multiple nanoantennas oriented in different directions. In general, a unit cell with $C_n$ rotational symmetry will exhibit an isotropic linear response if $n > 2$, with noncentrosymmetric configurations (odd $n$) required for net directional photocurrent. We demonstrate this with a Kagome lattice metasurface (Fig. 2b), which has three-fold rotational symmetry and three planes of mirror symmetry (*p3m1* wallpaper group). Global and corresponding local photocurrent directionality is continuously rotated by varying the incident linear polarization angle (Fig. 2d,e), with a linear combination of hot spot excitations (Fig. 2b) maintaining a constant photocurrent magnitude. This omni-directional current control is described in a Cartesian basis by $j_x = j \sum_i \cos^2(\theta - \theta_i) \cos(\theta_i)$ and $j_y = j \sum_i \cos^2(\theta - \theta_i) \sin(\theta_i)$, where $\theta$ is the laser polarization angle and $\theta_i = 0, 2\pi/3$, and $4\pi/3$ for the three nanoantenna orientations within the Kagome unit cell.

These prototypical examples already illustrate a broad design space for nanoscale vectorial photocurrents, which can be driven in arbitrary directions in space and/or time under different incident light fields. Localized vortical charge flows with no net global current can be realized under circularly-polarized excitation, for instance, by rotating the nanoantenna orientations within the Kagome metasurface unit cell (Fig. S11). The design space grows further with the introduction of different structures and resonant wavelengths. In such systems, the light-matter interaction symmetry may even be different than the underlying structural symmetry and tunable with wavelength, as read out by the polarization- and wavelength-dependent photocurrent response. Such possibilities are demonstrated in Extended Data Fig. 2, where the Kagome metasurface unit cell contains three resonators of different dimensions. Although this perturbed Kagome lattice evidently exhibits the lowest possible degree of symmetry (*p1* wallpaper group), an approximate mirror symmetry in the light-matter interaction appears either along a dominant resonance axis or between two equally excited resonances at intermediate wavelengths[24]. The general linear photocurrent response for any metasurface with uniform spatial patterning is given by $j_x(\theta, \lambda) \propto \sum_i E_i^2(\lambda) \cos^2(\theta - \theta_i) \cos(\theta_i)$ and $j_y(\theta, \lambda) \propto \sum_i E_i^2(\lambda) \cos^2(\theta - \theta_i) \sin(\theta_i)$, where $E_i(\lambda)$ are the wavelength-dependent plasmonic fields at each tip hot spot. This shows a high degree of adaptability, with responses that are not rigidly fixed by metasurface patterning. These results also directly demonstrate important capabilities for bias-free polarization-resolved (Fig. 2d,e) and wavelength-resolved (Extended Data Fig. 2) photodetection for integrated ultrafast optoelectronic and information science applications[37].

**Spatially-Varying Vectorial Photocurrents**

Nonuniform spatial patterning can be implemented for nearly arbitrary control over vectorial current distributions out to macroscopic scales, as demonstrated with radial (Fig. 3a) and azimuthal (Fig. 3e) metasurfaces. Corresponding radial and azimuthal transient charge flows generated upon uniform circularly-polarized excitation are expected to be evident in the far-field THz radiation patterns. We thus utilize a high-throughput THz emission spectroscopy system for hyperspectral imaging (Fig. 3d), scanning across the THz beam and reading out the full time, frequency, and polarization information as a function of position (Methods). These maps reveal clear radially-polarized (Fig. 3b,c) and azimuthally-polarized (Fig. 3f,g) THz fields, unambiguously confirming the vectorial photocurrents. For the radial metasurface device, the radial current and resonantly-enhanced excitation are also verified by direct photocurrent readout measurements (Extended Data Fig. 3).

While the mapped THz fields serve to demonstrate the spatially-varying vectorial currents, they also directly illustrate a new capability for direct generation of broadband vector beams in the THz frequency range (out to ~4 THz; Fig. 3d). These cylindrical THz vector beams hold significant promise for THz imaging, spectroscopy, and other applications[16]. Although various schemes have emerged recently for generating THz vector beams[16], they often rely on bulky, narrowband, and/or multi-stage systems. By contrast, direct generation of arbitrary THz vector fields from spatially-patterned photocurrents offers an ultrathin, broadband, and single-stage source that can also be actively manipulated by incident optical fields or by electrostatic gating, as examined next.

**Photothermoelectric Dynamics**

We now investigate the rich photothermoelectric dynamics that occur within the gold-graphene metasurfaces under femtosecond optical excitation at the nanometric tip hot spots. The photocurrent driving mechanism is revealed by electrostatic gating of large-area metasurface devices (Methods), with a back-gate voltage $V_g$ applied to tune chemical potential $\mu$ and corresponding electrical conductivity $\sigma$ in the bare graphene regions (Fig. 4a). By contrast, the chemical potential of the graphene beneath the nanoantennas remains pinned to that of gold[38], leading to a spatially-varying Seebeck coefficient (Fig. 4c) described by the Mott relation at room temperature $T_0$, $S_0 = -\frac{\pi^2 k_B^2 T_0}{3e} \frac{1}{\sigma} \frac{d\sigma}{d\mu}$, where $k_B$ is the Boltzmann constant, $e$ is the elementary charge, and $|\mu| \propto |V_g - V_{CNP}|^{1/2}$ with $V_{CNP} \approx 9$ V at the charge neutrality point (Supplementary Note 1). The photocurrent closely follows the nonmonotonic dependence of the difference $\Delta S_0$ between the bare and gold-doped graphene regions (Fig. 4b), which is a clear signature of a photothermoelectric effect[39,40] and is inconsistent with photovoltaic (monotonic $V_g$ dependence) or other linear responses that can occur at graphene junctions[14,33,34].

The dynamics studied here are extremely confined in both space and time, compared with previous investigations under continuous-wave excitation or at mesoscopic junctions[25,35,36,39-42]. We therefore combine electromagnetic, thermodynamic, and hydrodynamic modeling to understand the evolution of the far-from-equilibrium electronic temperature $T_e(\mathbf{r}, t)$, the effective photothermoelectric driving field $\propto -S(\mathbf{r}, t)\nabla T_e$, and the resulting nanoscale charge flow $\mathbf{j}(\mathbf{r}, t)$. The localized optical power absorption generates a peak $T_e$ approaching 3000 K at the nanoantenna tip (Fig. 4d) for an incident pulse fluence of 0.5 µJ cm$^{-2}$ in the linear response regime. In-plane electronic thermal diffusion and scattering with optical phonons[29] along with out-of-plane scattering with substrate phonons[30] then cool the superheated carrier distribution within a few hundred femtoseconds (Supplementary Note 3). Before cooling, a net acceleration proportional to $\Delta S \nabla T_e$ acts on the electronic system along the nanoantenna axis (Fig. 4e). We note that while $\Delta S$ exhibits a nonlinear temperature dependence at high $T_e$ (Supplementary Note 3), this is not expected to alter the gating behavior shown for $\Delta S_0$ (Fig. 4b) beyond a scaling factor. The resulting charge flow is modeled via the time-dependent incompressible Navier-Stokes equation (Fig. 4f; Methods). Such a hydrodynamic framework[43-45] is based on the condition that $\tau_{ee} \ll \tau_{\mathrm{mr}}$, where $\tau_{ee}$ is the electron-electron scattering time and $\tau_{\mathrm{mr}}$ is momentum relaxation time associated with electron-impurity/phonon scattering. This treatment can be justified even for relatively low-mobility graphene ($\tau_{\mathrm{mr}} \approx 45$ fs) by the electronic superheating, which dramatically increases the phase space for electron-electron scattering. Upon photoexcitation, $\tau_{ee} \propto T_e^{-2}$ is driven from a few hundred femtoseconds at room temperature into the few-femtosecond range at several thousand Kelvin (Supplementary Note 3), supporting a hydrodynamic treatment and the possibility of light-induced viscous electron flow[46].

Indeed, our multiphysics modeling reveals the rapid evolution of $T_e$, $\tau_{ee}$, and corresponding local physical quantities (e.g., electron viscosity, thermopower, and heat capacity) that occurs over hundreds-of-femtosecond timescales and tens-of-nanometer spatial scales (Extended Data Fig. 4). The calculations also capture the essential macroscopic behaviors measured for the various metasurfaces, including polarization, frequency, and intensity dependencies, with time-averaged photocurrents that agree to within a factor of 2 of experimental measurements (Fig. S9).

These robust model results elucidate the ultrafast nanoscale charge flows that produce net local and global vectorial photocurrents measured via THz emission and electrical readout. Beyond gold and graphene, the choice of materials for vectorial optoelectronic metasurfaces is expansive, with photocurrent responses expected due to photovoltaic, photoelectric, and photothermoelectric effects in semiconductor, semimetal, topological insulator, ferromagnetic and other material systems. Symmetry-broken nanoscale structuring may therefore also be imposed on new materials to probe unknown local responses and physical properties[3]. Furthermore, the introduction of planar chirality and thus net vorticity can lead to dynamic nano-magnetic moments (Supplementary Note 4). More generally, spatially-varying photocurrents can be utilized to generate complex magnetic fields[47] in free space or nearby materials. While we have directly demonstrated polarization/wavelength-resolved photodetection and ultrafast THz generation in symmetry-broken optoelectronic metasurfaces towards information science, microelectronics, and THz technologies, we anticipate that many more opportunities will emerge in other applications (e.g., nano-magnetism, materials diagnostics, analog computing, and energy harvesting) using the basic concepts introduced in this work.

## Methods

**Metasurface Fabrication**

Metasurface fabrication begins with large-area monolayer graphene grown via chemical vapor deposition (CVD) and transferred onto fused quartz substrates (Supplementary Note 1). Substrates are spin-coated with bilayer PMMA (PMMA 495 bottom layer and PMMA 950 top layer) at 3000 rpm for 30 s, followed by 1 min prebake at 180 °C. A conductive polymer layer (DisCharge) is then applied to ensure adequate charge dissipation during electron beam exposure. Arrays are then patterned via electron beam lithography using a 100 kV JEOL JBX-6300FS system. For < 30 nm features, interparticle proximity effect correction is utilized with an optimized total dosage around 1650 µC cm$^{-2}$. Under these conditions, writing of a 1 mm$^2$ array takes approximately 1 hr. Following a 30 s rinse in isopropanol (IPA) for DisCharge removal, samples are then developed in a 3:1 IPA:MIBK (methyl isobutyl ketone) solution for 1 min. A 30 nm gold layer is then deposited via electron-beam or sputter deposition without any adhesion layer. Liftoff is performed after overnight acetone soaking, using gentle acetone and IPA wash bottle rinsing to remove the residual PMMA/gold. For most samples, no particle delamination is observed.

Devices for simultaneous ultrafast THz emission and average photocurrent electrical readout studies are prepared using maskless photolithography (Heidelberg MLA 150). We utilize AZ 5214E photoresist in image reversal mode, precoating the samples with a hexamethyldisilazane (HMDS) layer to promote photoresist adhesion. Excess graphene is etched to obtain desired geometries via 2 min exposure to $O_2$ plasma (100 W, 10 sccm; Anatech RIE). Electrodes consist of 50 nm Au with a 3 nm Ti adhesion layer.

For back-gated devices, the back-gate electrode with minimal overlap with the top electrodes (precluding any shorting through defects in the dielectric spacer layer) is prepared via photolithographic patterning, deposition of 3 nm/30 nm Ti/Pt, and a liftoff process. Various spacer layer materials ($SiO_2$, $Al_2O_3$, and $HfO_2$), deposition methods (atomic layer deposition and physical vapor deposition), and thicknesses (5–100 nm) are tested for optimal electrostatic gating of large-area devices. We find that 30 nm $SiO_2$ offers a good trade-off between high gating capacitance and relatively modest nanoantenna resonance red-shifting due to the image charge oscillation within the Pt and dielectric environment of the spacer layer. Subsequent device preparation follows as described above.

**Basic Characterization**

The quality of metasurface fabrication is characterized via optical microscopy, scanning electron microscopy (SEM), and white light transmission spectroscopy. Optical bright and dark-field imaging under 50× magnification is sufficient for resolving individual nanostructures to verify successful liftoff and large-scale sample uniformity, while also resolving the monolayer graphene edge in etched devices. A more detailed view of the graphene and nanostructure morphologies (Figs. S1 and S2) is provided via SEM micrographs collected using FEI Magellan, Nova NanoSEM 450, and Nova NanoLab 600 systems. The fabricated nanoantennas closely reproduce the design profile down to the tens-of-nanometer scale, with extra-sharp ~15 nm radius of curvature tips.

Metasurface optical properties are verified via white light transmission spectroscopy using a tungsten-halogen white light source passed through a broadband polarizer for linear polarization control. Two 20× microscope objectives are utilized to focus the light through the sample and collect the transmitted light into both visible (Acton SP2300) and near-infrared (Ocean Optics NIRQuest) spectrometers. The transmittance is given by the ratio of signal (on the metasurface) to reference (off the metasurface but on the same substrate), with the ambient background collected with the white light source turned off and subtracted from each.

**Terahertz Emission Spectroscopy**

THz emission experiments are performed across several systems with different functionality, with overlapping datasets verifying reproducibility. All measurements except wavelength-dependent THz emission (see below) are performed in the low-fluence regime (< 1 μJ cm$^{-2}$), utilizing a Ti:sapphire oscillator (Chameleon Vision-S) providing 800 nm, ~100 fs pulses at the sample location with 80 MHz repetition rate. As seen in Fig. 1e, these low fluences ensure linear responses and also preclude thermally-induced nanoantenna deformation or photochemical degradative effects[48] that can occur in the intense hot spot regions. The gradual onset of a sub-linear regime beyond ~0.8 μJ cm$^{-2}$ in Fig. 1e is attributed to competing temperature-dependent thermodynamic contributions (Supplementary Notes 3 and 4). For metasurfaces operating at 800 nm, The THz radiation is measured via electro-optic sampling[49] using a 1 mm ⟨110⟩ ZnTe crystal. Three wire-grid polarizers are utilized to measure the $x$ and $y$ field components of the THz radiation, with the first polarizer at 0° or 90°, second polarizer at 45°, and third polarizer at 0°. The THz beam path is contained within a dry-air purged environment (< 2% relative humidity).

Pump-wavelength-dependent THz emission experiments are performed using a noncollinear optical parametric amplifier to generate pump pulses across 725–875 nm (pulse width ≲ 50 fs) and an optical parametric amplifier to generate pump pulses across 1450–1650 nm (pulse width ≲ 200 fs). Both amplifiers are seeded by a 1 MHz fiber amplified solid state laser producing ~270 fs pulses at 1032 nm. In all of these measurements, we tune the pump wavelength while the incident fluence is maintained at a constant value using a neutral density attenuator. The direct output of the seed laser is used to gate the THz emission using electro-optic sampling with a 0.5 mm thick ⟨110⟩ GaP crystal. This enables wavelength-dependent pumping without affecting the detection. This system is also utilized to obtain the THz emission for the 1550 nm metasurface presented in Fig. 1d, right panel.

For spatial mapping of THz vector beams, we utilize a custom-developed Menlo Systems apparatus, consisting of a 100 MHz mode-locked 1560 nm erbium-doped all-fiber laser oscillator that seeds two separate amplifiers. The first line uses a high-power erbium fiber amplifier driven in the linear pulse propagation regime followed by free-space second harmonic generation, outputting 1 W of 780 nm, 125 fs pulses and serving as the pump beam for generation of THz radiation from metasurfaces. The second amplifier line is used to gate a broadband, fiber-coupled PCA detector[50] (TERA 15-RX-PC), which is mounted on a two-dimensional stage for automated *xy* spatial scanning. Along with the high-repetition integrated delay stage (> 45 Hz for a 20 ps temporal scanning range), this system enables rapid hyperspectral imaging of THz vector beams collimated and focused by two TPX lenses (50 mm focal length). The PCA is situated before the focal point of the second lens, with resulting deviations from a planar phase-front calibrated and corrected for in the results presented in Fig. 3.

**Device Gating and Photocurrent Readout**

Time-averaged photocurrents are electrically read out via a lock-in amplifier (LIA; for the highest signal-to-noise ratio) and picoammeter (to directly measure polarity). Gating studies are performed using a computer-controlled dual-channel sourcemeter (Keithley 2614B) for simultaneous application of a gate voltage and readout of either the device resistance (via sourcemeter) or photocurrent (via picoammeter or LIA). While nonlocal photocurrents on electrodes are often evaluated in terms of a Shockley-Ramo type response[1,51], the square metasurface and electrode geometries here lead to a simplified, direct readout of the overall *x* and *y* photocurrent components. Wavelength-dependent photocurrent responses are measured by directly tuning the output wavelength of the 80 MHz Ti:sapphire laser while maintaining a constant power. While the electrode gold-graphene junctions can also contribute to the photothermoelectric current, the current induced by the nanoantennas is isolated via (i) the spatial position of the focused beam, (ii) the much stronger nanoantenna polarization dependence, and (iii) the lack of a similar response in nanoantenna-free devices (see Extended Data Fig. 5 for further details).

**Dynamical Multiphysics Model**

Coupled electromagnetic, thermodynamic, and hydrodynamic equations are solved using the finite element method (*COMSOL Multiphysics 6.1*). Nanoscale optical fields are first simulated using a 3D rectangular metasurface unit cell domain consisting of the air superstrate, gold nanoantennas (dielectric function given by Johnson and Christy[52]), conductive graphene boundary layer, and fused quartz substrate[53]. Periodic (Floquet) boundary conditions are applied in the transverse *x* and *y* directions, with input/output ports along the *z* direction backed by perfectly matched layers. The graphene optical conductivity is approximately constant in the near-infrared frequency range[54], with real component $\sigma_r = 6.1 \times 10^{-5}$ $\Omega^{-1}$ and imaginary component $\sigma_i = -2.1 \times 10^{-5}$ $\Omega^{-1}$. Resonant transmission spectra are obtained from the simulated S-parameters under plane wave excitation at normal incidence, consistent with the experiments. The same nanoantenna geometry file is used for both lithography and simulations.

The full thermal evolution of the 2D graphene electron and lattice (specifically, optical phonon[29]) systems is calculated through the coupled heat equations,

$$c_e \frac{\partial T_e}{\partial t} = \tfrac{1}{2}\sigma_r E^2 + \nabla \cdot (\kappa_e \nabla T_e) - g_{\mathrm{er}}(T_e - T_{\mathrm{op}}) - g_{\mathrm{sub}}(T_e - T_0), \qquad (1\mathrm{a})$$

$$c_{\mathrm{op}} \frac{\partial T_{\mathrm{op}}}{\partial t} = \kappa_l \nabla^2 T_{\mathrm{op}} + g_{\mathrm{er}}(T_e - T_{\mathrm{op}}), \qquad (1\mathrm{b})$$

where $c_e$ is the volumetric electronic specific heat described in Supplementary Note 3, $\kappa_e = \tfrac{1}{3}\left(\tfrac{\pi k_B}{e}\right)^2 T_e \sigma$ is the in-plane electronic thermal conductivity approximating Wiedemann-Franz law behavior (approximately preserved in the Fermi liquid regime of graphene[45]), $g_{\mathrm{sub}} = 3$ MW m$^{-2}$ K$^{-1}$ is utilized as the out-of-plane electronic thermal conductance due to coupling with the SiO$_2$ substrate phonons[30] (corresponding to a cooling length $\sqrt{\kappa_e/g_{\text{sub}}} \approx 150$ nm at 3000 K), $T_{\text{op}}$ is the optical phonon temperature, $c_{\text{op}}$ is the optical phonon specific heat based on time-resolved Raman studies of graphite[29,55], and $\kappa_l \approx 1.7 \times 10^{-7}$ W K$^{-1}$ is the lattice thermal conductance based on thermal transport measurements of supported graphene[56] (1:1 device aspect ratio, here). The energy relaxation electron–optical phonon coupling constant, $g_{\text{er}} \approx 2 \times 10^7$ W K$^{-1}$m$^{-2}$, is estimated from transient reflectivity measurements (Supplementary Note 3). The source term, $\frac{1}{2}\sigma_r E^2$, is the absorbed power density from the transient laser pulse field $E(t) = E_0 e^{-2\ln 2\, t^2/\tau_p^2}$, where $E_0$ is the peak field determined in the frequency-domain electromagnetic simulations and $\tau_p = 100$ fs is the laser pulse duration. Coupling with the acoustic phonon bath on longer picosecond timescales and radiative loss are neglected. The effect of Peltier cooling—which would appear as $-T_e \nabla S \cdot \mathbf{j}$ on the right-hand side of the heat equation—is found to be small. Further details on the thermodynamic modeling are provided in Supplementary Note 3, including the thermodynamic quantities in the high-$T_e$ limit in graphene and simplified two-temperature models for the electron and lattice temperature evolution.

The electronic temperature evolution (energy flow) described by Eq. 1 drives momentum flow via photothermoelectric force $\mathbf{f}_{\text{PTE}} \propto \nabla T_e$, as modeled by the time-dependent linearized Navier-Stokes equation,

$$\frac{\partial \mathbf{u}}{\partial t} - \nabla \cdot (\nu \nabla \mathbf{u}) + \frac{1}{\tau_{\text{mr}}}\mathbf{u} = \mathbf{f}_{\text{PTE}}, \qquad (2)$$

assuming incompressible flow of the charged fluid,

$$\nabla \cdot \mathbf{u} = 0.$$

In these hydrodynamic equations, $\mathbf{u}$ is the 2D electron velocity field (current density $\mathbf{j} = e n_e \mathbf{u}$ with $n_e$ the charge density) and $\nu$ is the temperature-dependent kinematic electron viscosity. As is typical for electron hydrodynamic flow, the Reynolds number is small and the contribution of the nonlinear convective term $(\mathbf{u} \cdot \nabla \mathbf{u})$ is found to be negligible. The electron viscosity depends linearly on the electron-electron scattering time[43], $\nu \propto \tau_{ee}$, which therefore scales locally as $\sim T_e^{-2}$ (Supplementary Note 3). The diffusive term is written as $\nabla \cdot (\nu \nabla \mathbf{u})$, similar to the diffusive term in the heat equation, to properly account for the spatial variation of the kinematic viscosity (Extended Data Fig. 4). The temperature dependence of $S$ in the force term ($\mathbf{f}_{\text{PTE}} = -qS\nabla T_e$, where $q = \pm e$) is determined by solving the generalized Mott relation (Supplementary Note 3). Small contributions from Joule heating, which depends quadratically on the current density, are not included in the present treatment. The resulting ultrafast, nanoscale spatiotemporal evolution of the electronic temperature, force field, charge flow, and $\tau_{ee}/\tau_{\mathrm{mr}}$ are shown in Extended Data Fig. 4, with further discussion of the hydrodynamic modeling provided in Supplementary Note 4.

## Author Information

### Notes

The authors declare no competing interests.

### Contributions

Experiments were conceived by JP, HTC, RPP, and PP and experimental measurements were performed by JP, PP, TS, LG, LM, LY, YH, and PA, with input and resources provided by RH, AKA, AJT, FR, RPP, and HTC. Nanofabrication was performed by JP, CCC, JN, KWCK, and JKB with input from AKA and HTC. Numerical models were developed by JP and SZL. The manuscript was prepared by JP and HTC, with contributions and final approval from all authors.

## Acknowledgements

The authors thank Nicholas Sirica, Anatoly V. Efimov, Jinkyoung Yoo, and Ting S. Luk for helpful conversations, as well as Anthony James and Denise B. Webb for support on the nanofabrication process development. Partial support for this work was provided by the Los Alamos National Laboratory Laboratory-Directed Research and Development (LDRD) program. KWCK acknowledges support from the DOE National Nuclear Security Administration (NNSA) Laboratory Residency Graduate Fellowship Program No. DE-NA0003960. FR was supported by the US DOE Basic Energy Sciences, Division of Materials Science and Engineering project 'Quantum Fluctuations in Narrow-Band Systems'. Work was primarily performed at the Center for Integrated Nanotechnologies, an Office of Science User Facility operated for the U.S. Department of Energy (DOE) Office of Science. Los Alamos National Laboratory, an affirmative action equal opportunity employer, is managed by Triad National Security, LLC for the U.S. Department of Energy NNSA, under Contract No. 89233218CNA000001.

# Data Availability

All data and calculations available upon reasonable request.

# References

1   Ma, Q., Krishna Kumar, R., Xu, S.-Y., Koppens, F. H. L. & Song, J. C. W. Photocurrent as a multiphysics diagnostic of quantum materials. *Nat. Rev. Phys.* **5**, 170-184 (2023).

2   Orenstein, J. *et al.* Topology and symmetry of quantum materials via nonlinear optical responses. *Annu. Rev. Condens. Matter Phys.* **12**, 247-272 (2021).

3   Pettine, J. *et al.* Ultrafast terahertz emission from emerging symmetry-broken materials. *Light Sci. Appl.* **12**, 133 (2023).

4   Takasan, K., Morimoto, T., Orenstein, J. & Moore, J. E. Current-induced second harmonic generation in inversion-symmetric Dirac and Weyl semimetals. *Phys. Rev. B* **104**, L161202 (2021).

5   Sirica, N. *et al.* Photocurrent-driven transient symmetry breaking in the Weyl semimetal TaAs. *Nat. Mater.* **21**, 62-66 (2022).

6   Dupont, E., Corkum, P. B., Liu, H. C., Buchanan, M. & Wasilewski, Z. R. Phase-controlled currents in semiconductors. *Phys. Rev. Lett.* **74**, 3596-3599 (1995).

7   Schiffrin, A. *et al.* Optical-field-induced current in dielectrics. *Nature* **493**, 70-74 (2013).

8   Sederberg, S. *et al.* Vectorized optoelectronic control and metrology in a semiconductor. *Nat. Photonics* **14**, 680-685 (2020).

9   Boolakee, T. *et al.* Light-field control of real and virtual charge carriers. *Nature* **605**, 251-255 (2022).

10  McIver, J. W., Hsieh, D., Steinberg, H., Jarillo-Herrero, P. & Gedik, N. Control over topological insulator photocurrents with light polarization. *Nat. Nanotechnol.* **7**, 96-100 (2012).

11  Wang, Y. X. *et al.* Visualization of bulk and edge photocurrent flow in anisotropic Weyl semimetals. *Nat. Phys.* **19**, 507-514 (2023).

12  Kampfrath, T. *et al.* Terahertz spin current pulses controlled by magnetic heterostructures. *Nat. Nanotechnol.* **8**, 256-260 (2013).

13  Qiu, H. S. *et al.* Ultrafast spin current generated from an antiferromagnet. *Nat. Phys.* **17**, 388-394 (2021).

14  Koppens, F. H. L. *et al.* Photodetectors based on graphene, other two-dimensional materials and hybrid systems. *Nat. Nanotechnol.* **9**, 780-793 (2014).

15  Schubert, O. *et al.* Sub-cycle control of terahertz high-harmonic generation by dynamical Bloch oscillations. *Nat. Photonics* **8**, 119-123 (2014).

16  Petrov, N. V., Sokolenko, B., Kulya, M. S., Gorodetsky, A. & Chernykh, A. V. Design of broadband terahertz vector and vortex beams: I. Review of materials and components. *Light Adv. Manuf.* **3**, 1-19 (2022).

17  Mubeen, S. *et al.* An autonomous photosynthetic device in which all charge carriers derive from surface plasmons. *Nat. Nanotechnol.* **8**, 247-251 (2013).

18  Müller, M., Paarmann, A. & Ernstorfer, R. Femtosecond electrons probing currents and atomic structure in nanomaterials. *Nat. Commun.* **5**, 5292 (2014).

19  Ma, E. Y. *et al.* Recording interfacial currents on the subnanometer length and femtosecond time scale by terahertz emission. *Sci. Adv.* **5**, eaau0073 (2019).

20  Linic, S., Chavez, S. & Elias, R. Flow and extraction of energy and charge carriers in hybrid plasmonic nanostructures. *Nat. Mater.* **20**, 916-924 (2021).

21  Pettine, J. & Nesbitt, D. J. Emerging methods for controlling hot carrier excitation and emission distributions in nanoplasmonic systems. *J. Phys. Chem. C* **126**, 14767-14780 (2022).

22  Dombi, P. *et al.* Ultrafast strong-field photoemission from plasmonic nanoparticles. *Nano Lett.* **13**, 674-678 (2013).

23  Lehr, M. *et al.* Momentum distribution of electrons emitted from resonantly excited individual gold nanorods. *Nano Lett.* **17**, 6606-6612 (2017).

24  Pettine, J., Choo, P., Medeghini, F., Odom, T. W. & Nesbitt, D. J. Plasmonic nanostar photocathodes for optically-controlled directional currents. *Nat. Commun.* **11**, 1367 (2020).

25  Wei, J. X. *et al.* Zero-bias mid-infrared graphene photodetectors with bulk photoresponse and calibration-free polarization detection. *Nat. Commun.* **11**, 6404 (2020).

26  Wei, J., Xu, C., Dong, B., Qiu, C.-W. & Lee, C. Mid-infrared semimetal polarization detectors with configurable polarity transition. *Nat. Photonics* **15**, 614-621 (2021).

27  Li, L. F. *et al.* Room-temperature valleytronic transistor. *Nat. Nanotechnol.* **15**, 743-749 (2020).

28  Liao, P. F. & Wokaun, A. Lightning rod effect in surface enhanced Raman scattering. *J. Chem. Phys.* **76**, 751-752 (1982).

29  Lui, C. H., Mak, K. F., Shan, J. & Heinz, T. F. Ultrafast photoluminescence from graphene. *Phys. Rev. Lett.* **105**, 127404 (2010).

30  Low, T., Perebeinos, V., Kim, R., Freitag, M. & Avouris, P. Cooling of photoexcited carriers in graphene by internal and substrate phonons. *Phys. Rev. B* **86**, 045413 (2012).

31  Johannsen, J. C. *et al.* Direct view of hot carrier dynamics in graphene. *Phys. Rev. Lett.* **111**, 027403 (2013).

32  Luo, L. *et al.* Broadband terahertz generation from metamaterials. *Nat. Commun.* **5**, 3055 (2014).

33  Mueller, T., Xia, F., Freitag, M., Tsang, J. & Avouris, P. Role of contacts in graphene transistors: A scanning photocurrent study. *Phys. Rev. B* **79**, 245430 (2009).

34  Liu, C. H. *et al.* Ultrafast lateral photo-Dember effect in graphene induced by nonequilibrium hot carrier dynamics. *Nano Lett.* **15**, 4234-4239 (2015).

35  Yoshioka, K. *et al.* Ultrafast intrinsic optical-to-electrical conversion dynamics in a graphene photodetector. *Nat. Photonics* **16**, 718-723 (2022).

36  Tielrooij, K. J. *et al.* Hot-carrier photocurrent effects at graphene-metal interfaces. *J. Phys. Condens. Mat.* **27**, 164207 (2015).

37  Mueller, T., Xia, F. N. A. & Avouris, P. Graphene photodetectors for high-speed optical communications. *Nat. Photonics* **4**, 297-301 (2010).

38  Giovannetti, G. *et al.* Doping graphene with metal contacts. *Phys. Rev. Lett.* **101**, 026803 (2008).

39  Gabor, N. M. *et al.* Hot carrier-assisted intrinsic photoresponse in graphene. *Science* **334**, 648-652 (2011).

40  Shautsova, V. *et al.* Plasmon induced thermoelectric effect in graphene. *Nat. Commun.* **9**, 5190 (2018).

41  Xu, X. D., Gabor, N. M., Alden, J. S., van der Zande, A. M. & McEuen, P. L. Photo-thermoelectric effect at a graphene interface junction. *Nano Lett.* **10**, 562-566 (2010).

42  Echtermeyer, T. J. *et al.* Photothermoelectric and photoelectric contributions to light detection in metal-graphene-metal photodetectors. *Nano Lett.* **14**, 3733-3742 (2014).

43  Levitov, L. & Falkovich, G. Electron viscosity, current vortices and negative nonlocal resistance in graphene. *Nat. Phys.* **12**, 672-676 (2016).

44  Bandurin, D. A. *et al.* Negative local resistance caused by viscous electron backflow in graphene. *Science* **351**, 1055-1058 (2016).

45  Crossno, J. *et al.* Observation of the Dirac fluid and the breakdown of the Wiedemann-Franz law in graphene. *Science* **351**, 1058-1061 (2016).

46  Block, A. *et al.* Observation of giant and tunable thermal diffusivity of a Dirac fluid at room temperature. *Nat. Nanotechnol.* **16**, 1195-1200 (2021).

47  Jana, K. *et al.* Reconfigurable electronic circuits for magnetic fields controlled by structured light. *Nat. Photonics* **15**, 622-627 (2021).

48  Mitoma, N., Nouchi, R. & Tanigaki, K. Photo-oxidation of graphene in the presence of water. *J. Phys. Chem. C* **117**, 1453-1456 (2013).

49  Nahata, A., Weling, A. S. & Heinz, T. F. A wideband coherent terahertz spectroscopy system using optical rectification and electro-optic sampling. *Appl. Phys. Lett.* **69**, 2321-2323 (1996).

50  Kohlhaas, R. B. *et al.* Ultrabroadband terahertz time-domain spectroscopy using III-V photoconductive membranes on silicon. *Opt. Express* **30**, 23896-23908 (2022).

51  Song, J. C. W. & Levitov, L. S. Shockley-Ramo theorem and long-range photocurrent response in gapless materials. *Phys. Rev. B* **90**, 075415 (2014).

52   Johnson, P. B. & Christy, R. W. Optical constants of noble metals. *Phys. Rev. B* **6**, 4370-4379 (1972).

53   Malitson, I. H. Interspecimen comparison of refractive index of fused silica. *J. Opt. Soc. Am.* **54**, 1205-1209 (1965).

54   Chang, Y. C., Liu, C. H., Liu, C. H., Zhong, Z. H. & Norris, T. B. Extracting the complex optical conductivity of mono- and bilayer graphene by ellipsometry. *Appl. Phys. Lett.* **104**, 261909 (2014).

55   Yan, H. G. *et al.* Time-resolved Raman spectroscopy of optical phonons in graphite: Phonon anharmonic coupling and anomalous stiffening. *Phys. Rev. B* **80**, 121403(R) (2009).

56   Seol, J. H. *et al.* Two-dimensional phonon transport in supported graphene. *Science* **328**, 213-216 (2010).

# Figures

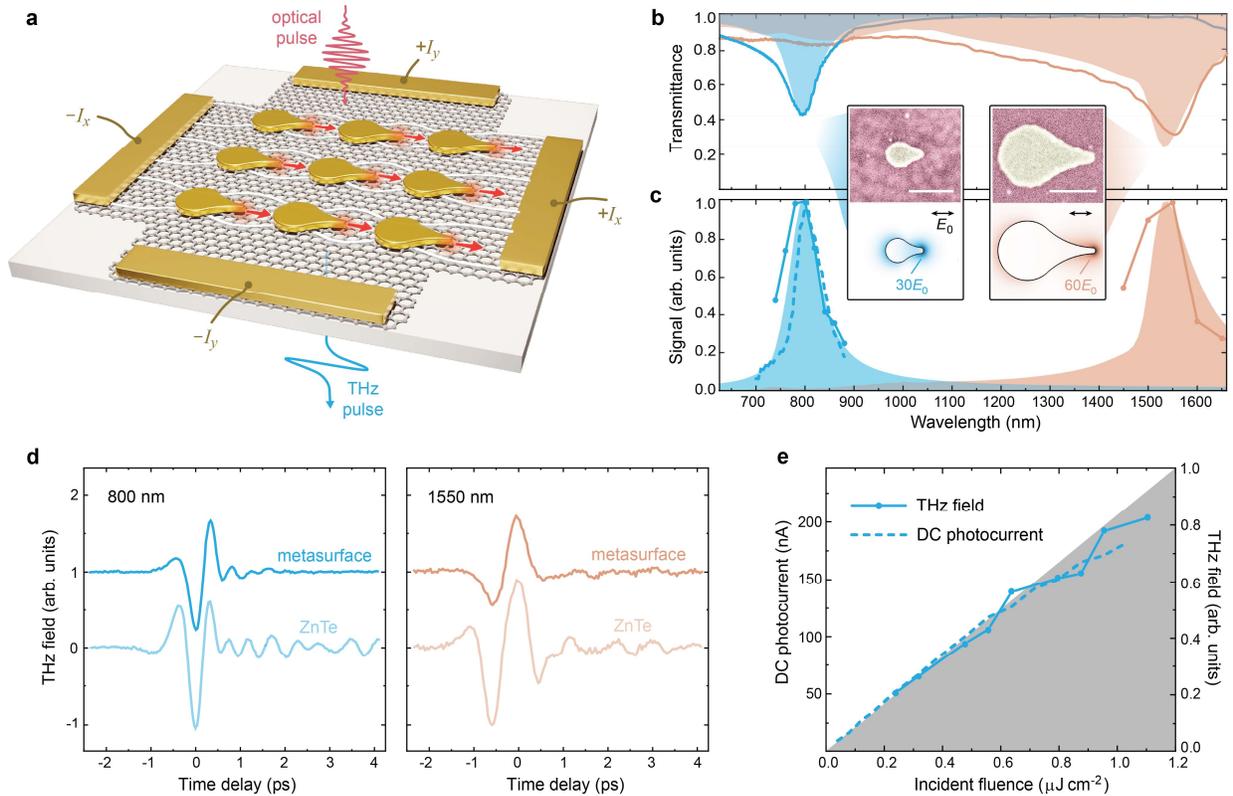

**Fig. 1 | Directional photocurrents in symmetry-broken optoelectronic metasurfaces. a**, Illustration of an optoelectronic metasurface consisting of symmetry-broken gold nanoantennas on graphene. Femtosecond laser illumination stimulates vectorial photocurrents and consequent emission of ultrafast THz pulses. **b**, Measured (solid line) and simulated (solid fill) transmission spectra for two nanoantenna designs with resonances at 800 nm and 1550 nm. **c**, Measured incident wavelength-dependent THz field amplitude (solid lines with data markers) and DC photocurrent (dashed line), as well as simulated field intensity (solid fills). Top insets: Scanning electron micrographs (SEMs) of the fabricated nanoantenna elements. Bottom insets: Simulated plasmonic field enhancements. Scale bars, 200 nm. **d**, Measured THz time-domain signals emitted from the resonantly-excited 800 nm (left) and 1550 nm (right) metasurfaces, compared with those from 1 mm ZnTe. Top curves are offset for clarity. **e**, Incident fluence-dependent THz field amplitude and DC photocurrent readout, along with a linear fit to low fluence (solid fill).

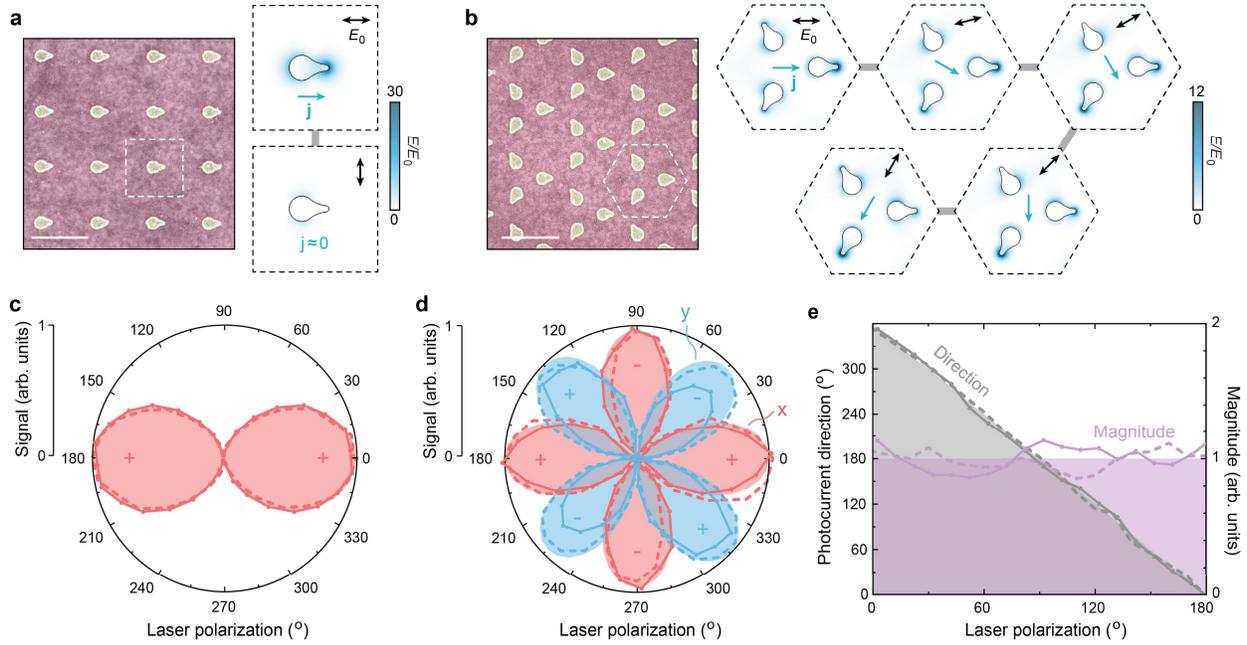

**Fig. 2 | Polarization-dependent local responses and omni-directional control. a,b**, SEM images of uniformly-oriented (**a**) and Kagome (**b**) metasurfaces with 800 nm resonances. Scale bars, 500 nm. Insets: Simulated resonant plasmonic field enhancements for different incident linear polarization angles (black double arrows), with the calculated net current direction indicated (blue single arrows). **c,d**, Measured *x* (red) and *y* (blue) components of the radiated THz field (solid lines with data markers) and photocurrent (dashed lines) for the uniformly-oriented (**c**) and Kagome (**d**) metasurfaces with respect to the incident linear polarization angle. Calculated linear responses (solid fills) are shown for comparison, with ± signs indicating lobe polarity. For clarity, a small residual *y* component is not shown in (**c**). **e**, The Kagome metasurface exhibits nearly constant photocurrent magnitude (purple) and continuously-rotatable direction (gray), consistent with analytic predictions.

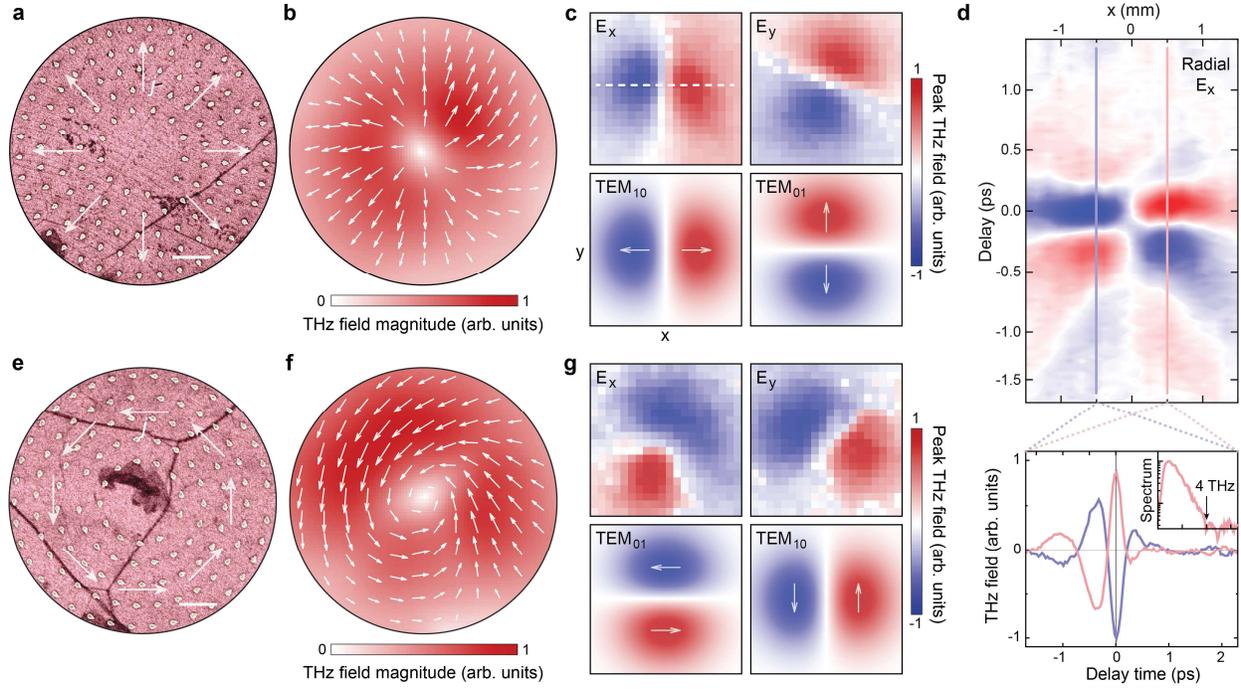

**Fig. 3 | Global vectorial currents and THz vector beams. a**, SEM image of the central region of a radial vector metasurface, with white arrows illustrating photocurrent in the radial direction upon circularly-polarized excitation. Scale bar, 1 μm. **b**, Spatial mapping of radial THz vector field. **c**, Measured (top) versus ideal (bottom) Hermite-Gaussian modes for the $x$ and $y$ field components of the radial THz vector beam. **d**, Lineout from the radial $E_x$ image, showing the transient THz field as a function of $x$ position, along with example THz time-domain waveforms illustrating the polarity reversal across the beam. Inset: The corresponding THz field amplitude spectrum in the frequency-domain on a logarithmic scale. **e–g**, same as (**a**–**c**) but for the azimuthal vector metasurface and THz beam.

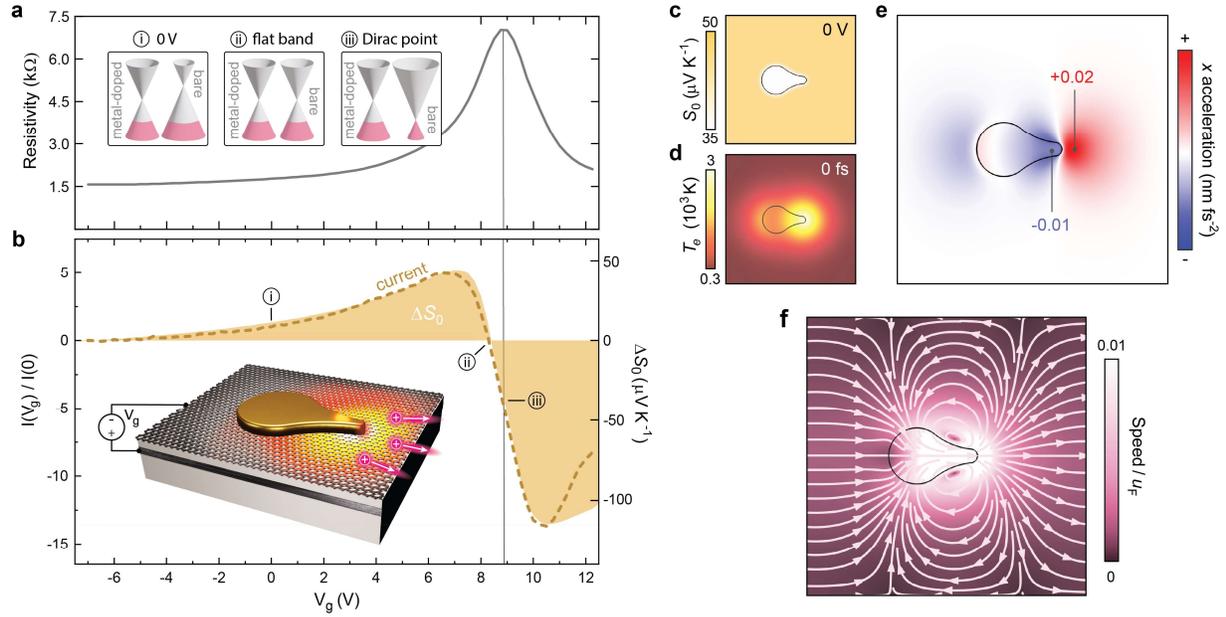

**Fig. 4 | Local photothermoelectric driving force and resulting nanoscale charge flow. a**, Measured resistivity of a metasurface device as a function of gate voltage. Insets: Chemical potentials of the gold-pinned and bare graphene regions for three gate voltages as indicated in (**b**). **b**, Measured gate-dependent current (dashed line) compared with calculated $\Delta S_0$ (solid fill). Inset: Illustration of the gated device configuration, with tip-localized heating and corresponding net carrier motion. **c**, Spatial distribution of $S_0$. **d**, Snapshot of the calculated $T_e$ distribution at the time of the incident pulse peak (0 fs). **e**, Photothermoelectric $x$ acceleration field (along the principal nanoantenna axis) at $V_g = 0$ V and 0 fs. **f**, Hydrodynamic velocity field for the hole fluid at $V_g = 0$ V and 0 fs.

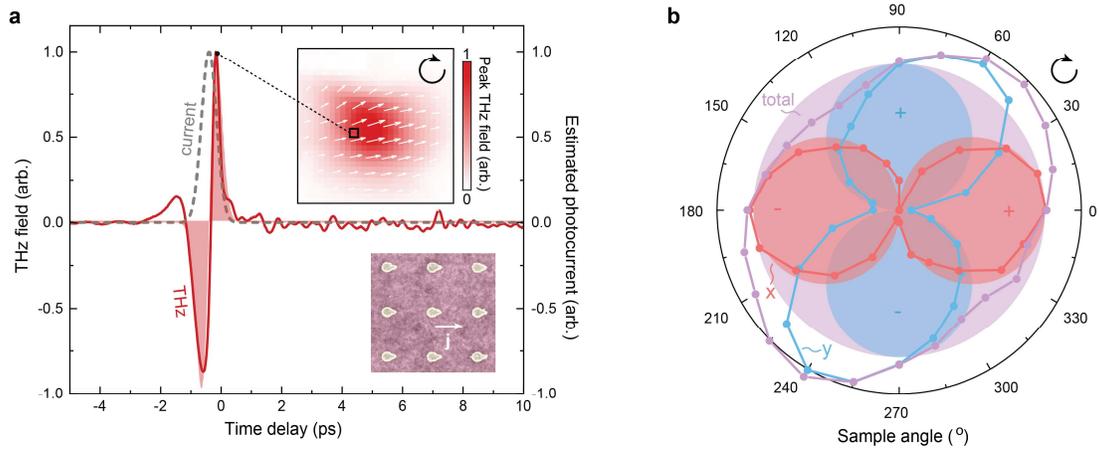

**Extended Data Fig. 1 | Linearly-polarized THz emission from directional photocurrent. a**, Measured (solid line) and approximated (solid fill) THz time-domain waveforms for a uniformly-oriented metasurface under circularly-polarized femtosecond laser excitation, assuming a Gaussian photocurrent pulse (gray dashed line; 600 fs full width at half maximum) with $E_{\text{THz}} \propto -dj/dt$. The true current and THz field is expected to be faster than estimated here due to bandwidth limitations of the photoconductive antenna detector. Inset: Spatially mapped THz field showing the beam with approximately uniform linear polarization. A small $y$-polarized contribution and corresponding ~10° angular deviation from $x$ axis may be attributed in part to residual sample tilt and imperfect alignment of the THz imaging system. **b**, The $x$ (red), $y$ (blue), and total (purple) THz field amplitudes measured (solid lines with data markers) as a function of sample orientation under circularly-polarized incident optical pulses, compared with the expected $\cos(\theta)$ (red fill), $\sin(\theta)$ (blue cyan fill), and constant (purple fill) behaviors, respectively. The skew in the $y$ dataset is attributed to imperfect circular polarization and residual misalignment in the parabolic mirrors.

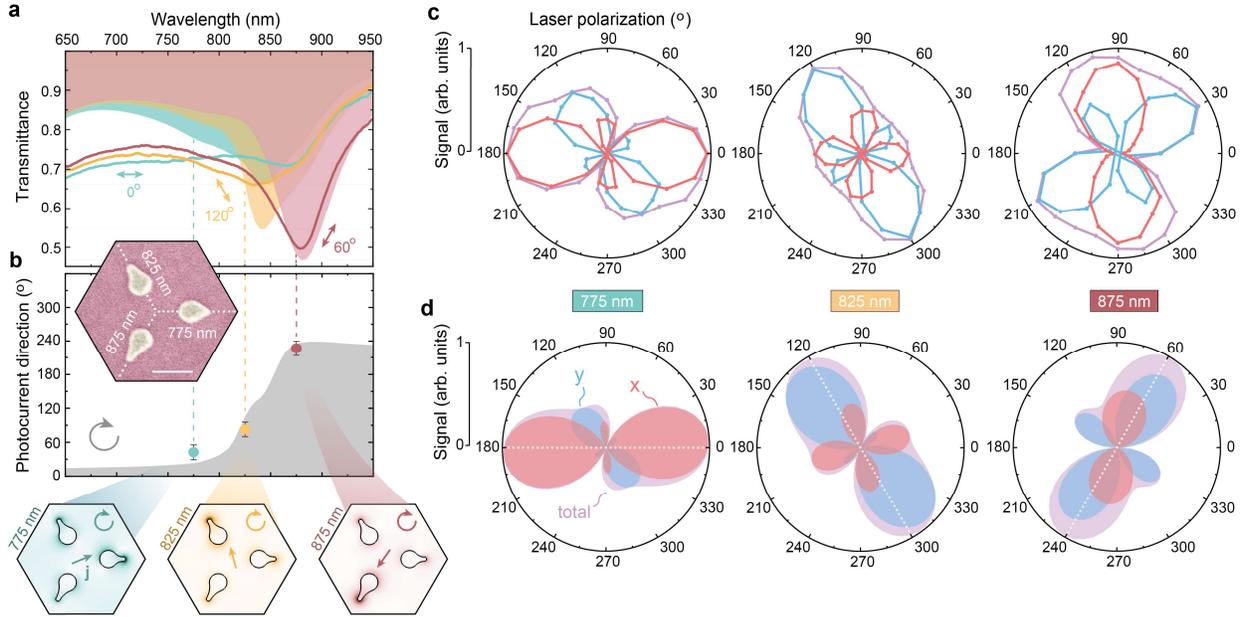

**Extended Data Fig. 2 | Wavelength-sensitive directional photocurrent response. a**, Measured (solid lines) and simulated (solid fills) transmittance spectra for the perturbed Kagome lattice, with white light linearly polarized along the three nanoantennas and resonances nominally designed around 775 nm (0°; cyan), 825 nm (120°; orange), and 875 nm (240°; red). Inset: Unit cell (scale bar, 250 nm). **b**, Direction of THz field polarization (and thus net photocurrent) measured (solid circles) and calculated (solid fills) as a function of incident wavelength for circularly-polarized incident light. Bottom insets show simulated fields and net current directions for circularly-polarized excitation at the three resonant wavelengths. **c**, Measured and **d**, calculated $x$ (red), $y$ (blue), and total (purple) THz fields as a function of incident linear polarization angle at the three resonance wavelengths, showing preferred alignment along the active resonance axes (white dotted lines).

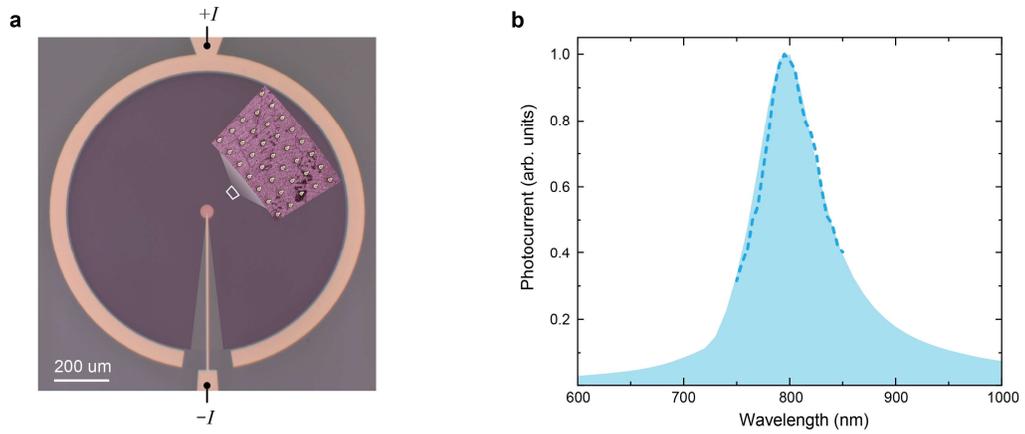

**Extended Data Fig. 3 | Radial metasurface device and resonant excitation. a**, Optical micrograph of a radial metasurface, prepared as a device for photocurrent readout. Inset: Radially-outward-oriented nanoantennas with resonance at 800 nm and nanoantenna spacing ~500 nm. **b**, Measured DC photocurrent (dashed line) and simulated plasmonic hot spot field intensity (solid fill; same as in Fig. 1c), verifying the resonantly-coupled/enhanced radial photocurrent.

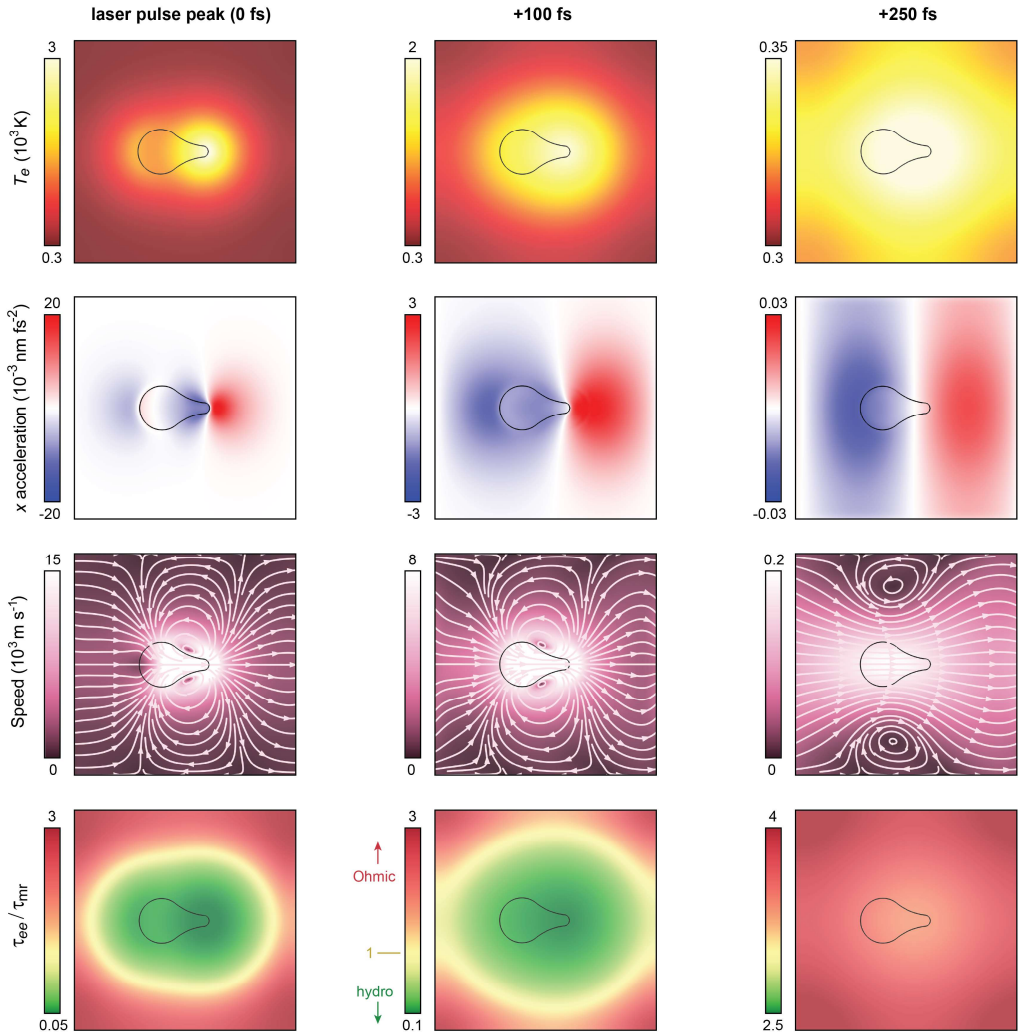

**Extended Data Fig. 4 | Ultrafast time evolution of local thermodynamic and hydrodynamic quantities.** Calculated electron temperature, photothermoelectric acceleration field, hydrodynamic flow profile, and relative scattering rates within the square metasurface unit cell, as a function of time with respect to the 100 fs laser pulse. All plots are normalized to the peak values.

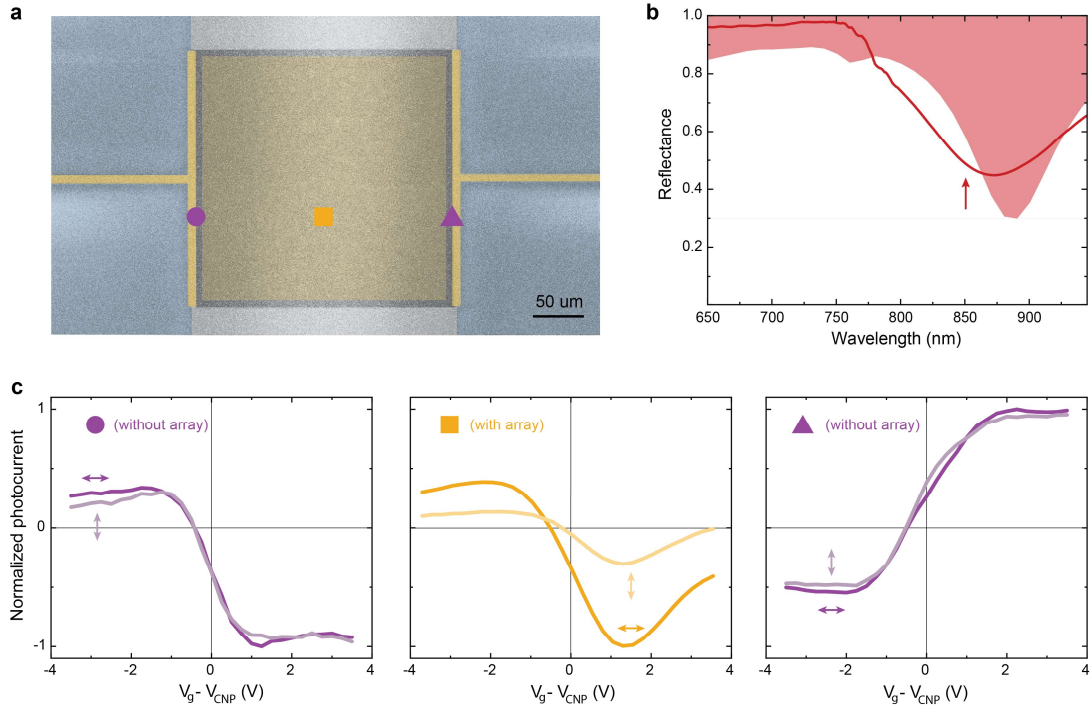

**Extended Data Fig. 5 | Electrostatically gated metasurface. a**, Metasurface device, with false coloring indicating the 3 nm/50 nm Ti/Au (light yellow), nanoantenna array (darker yellow), graphene (dark gray), 30 nm Pt back-gate electrode with 30 nm silica spacer layer (light gray), and substrate (light blue). **b**, Measured (solid line) and simulated (solid fill) reflectance spectrum of device, showing resonant absorption around 875 nm. **c**, Gate-dependent photocurrent as a function of position, with locations indicated in (**a**). Beam spot ~25 μm. To isolate the electrode contributions, the left and right electrode measurements are performed on an array-free device, with little laser polarization dependence observed. A polarity reversal is observed between the left and right electrodes. By contrast, strong polarization dependence is observed on the array at the center of the sample. Both metasurface and array-free devices have 250 μm square graphene between the electrodes (metasurface device is same as in Fig. 4). Photocurrents are normalized to the peak positive or negative values.

Supplementary Information for

# Light-Driven Nanoscale Vectorial Currents

Jacob Pettine[1]*, Prashant Padmanabhan[1], Teng Shi[1], Lauren Gingras[2], Luke McClintock[1,3], Chun-Chieh Chang[1], Kevin W. C. Kwock[1,4], Long Yuan[1], Yue Huang[1], John Nogan[5], Jon K. Baldwin[1], Peter Adel[2], Ronald Holzwarth[2], Abul K. Azad[1], Filip Ronning[6], Antoinette J. Taylor[1], Rohit P. Prasankumar[1,7], Shi-Zeng Lin[1], and Hou-Tong Chen[1]*

[1] Center for Integrated Nanotechnologies, Los Alamos National Laboratory, Los Alamos, NM 87545, United States

[2] Menlo Systems, Martinsried, Bavaria 82152, Germany

[3] Department of Physics, University of California-Davis, Davis, CA 95616, United States

[4] The Fu Foundation School of Engineering and Applied Science, Columbia University, New York, NY 10027, United States

[5] Center for Integrated Nanotechnologies, Sandia National Laboratories, Albuquerque, NM 87123, United States

[6] Institute for Material Science, Los Alamos National Laboratory, Los Alamos, NM 87545, United States

[7] Intellectual Ventures, Bellevue, WA 98005, United States

* jacob.pettine@lanl.gov, chenht@lanl.gov

## Contents

# Supplementary Note 1: Sample characterization

## 1.1 Monolayer graphene properties

Large-area (1 cm$^2$) graphene monolayers are obtained from commercial sources (Graphenea and ACS Material), with similar properties from both sources, grown via chemical vapor deposition (CVD) and transferred onto fused quartz substrates or gating devices by the standard wet transfer method using polymethyl methacrylate (PMMA)[1]. Water molecules trapped within the graphene/substrate interface during this process enhance the p-type graphene doping and can contribute to hysteresis under applied back-gate voltages ($V_g$) during electrostatic gating[2]. While hysteresis is observed within our electrostatic gating measurements, making the precise determination of the chemical potential more challenging, the photocurrents follow the measured resistivities in a consistent manner. We therefore utilize only the forward $V_g$ scans and corresponding photocurrents to gain physical insight on the photothermoelectric process.

Common microscopic domain structure for CVD graphene, including wrinkles and multilayer nucleation sites[1], are observed in scanning electron micrographs (Fig. S1). Although these defects contribute to momentum-relaxing scattering events and thereby limit global mobilities, little influence on the metasurface response is expected due to their small fill factor. Furthermore, only a small difference in the terahertz (THz) emission strength is observed for 1–4-layer samples, discussed below (Fig. S4). Longitudinal mobilities of $\mu_e \approx 1500$ cm$^2$V$^{-1}$s$^{-1}$ are determined by the resistivity of large-area ($\geq 200 \times 200$ μm$^2$) nano-patterned devices with negligible contact resistance. This is in good agreement with Hall mobilities measured on smaller Hall bar devices without nanostructures, indicating that residual resist and other factors from the lithography steps do not contribute significantly to scattering. Given these mobilities, we find $\tau_{\mathrm{mr}} = \frac{m^* \mu_e}{e} \approx 45$ fs, with effective mass $m^* = \frac{\hbar}{u_F}\sqrt{\pi |n_e|} = 0.05 m_e$ for the environmentally-doped monolayer graphene[3] (i.e., at zero gate voltage). The Fermi velocity is $u_F = 10^6$ m s$^{-1}$ and the measured charge (hole) density is $n_e = -\frac{\epsilon}{ed}(V_g - V_{\mathrm{CNP}}) = 6 \times 10^{12}$ cm$^{-2}$ at $V_g = 0$ for $V_{\text{CNP}} \approx 9$ V (see main text), where $\epsilon/\epsilon_0 = 3.9$ and $d = 30$ nm are the dielectric constant and thickness, respectively, of the SiO$_2$ spacer layer in the graphene devices. This corresponds to an environmentally-doped Fermi energy $\varepsilon_F = \pm \hbar u_F (\pi |n_e|)^{\frac{1}{2}} = -300$ meV at $V_g = 0$ for these samples.

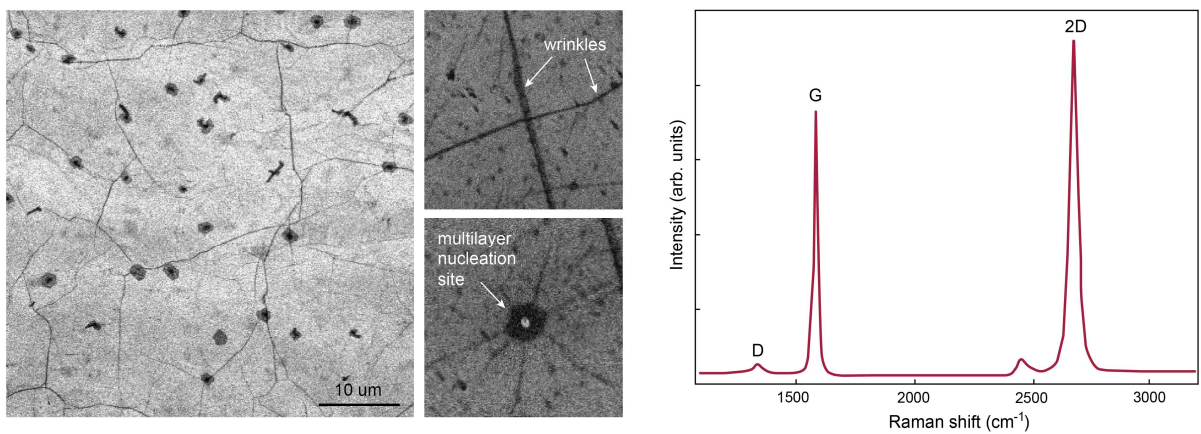

**Fig. S1 | Graphene properties.** (Left) Scanning electron micrographs showing typical features of large-area CVD graphene, including wrinkles and multilayer growth nucleation sites from which the monolayer domains grow. (Right) Raman spectrum of the monolayer graphene, with $I_{2D} > I_G \gg I_D$.

## 1.2 Nanoantenna structure and effect of exposed graphene removal

Gold nanostructures are fabricated with 15 nm radius of curvature tips, with good sticking between the gold and graphene obviating the need for a (dissipative) metal adhesion layer. To test the role of the graphene currents in THz generation, we remove the exposed graphene in a metasurface device via a minimally destructive 45 s plasma etch (100 W, 10 sccm O$_2$; Fig. S2a,b). This causes a small blueshift attributed to slight tip deformation during the plasma treatment (Fig.

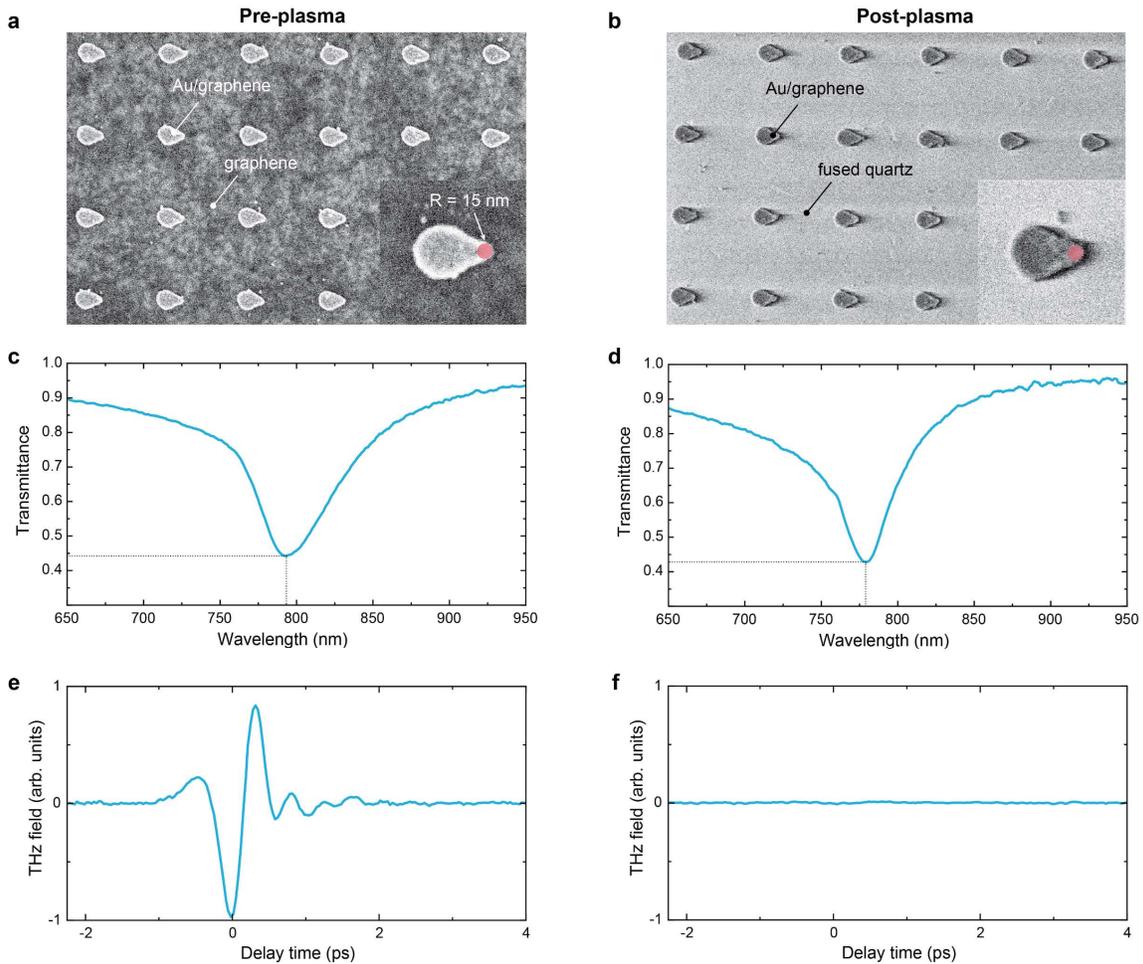

**Fig. S2 | Effect of bare graphene removal on metasurface resonance and THz emission. a**, Scanning electron micrograph before and, **b**, after removal of exposed graphene between nanoantennas. Insets: Individual nano-antennas with 15 nm radius circle illustrating tip radius. **c**, White light transmission spectrum before and, **d**, after graphene removal. **e**, THz time trace before and, **f**, after graphene removal.

S2c,d) but maintains the nanostructure integrity and strong plasmonic resonance (similar quality factor). Graphene can thus serve as an atomically-thin, minimally-damping adhesion layer between the gold and the underlying fused quartz substrate (or, in general, a variety of materials upon which graphene can be readily transferred). Most importantly for the present studies, the conductivity drops to zero after etching away the bare graphene and no photocurrents are observed. The disappearance of the THz radiation (Fig. S2e,f) under otherwise identical testing conditions illustrates the central role of the graphene photocurrents, as opposed to any local currents are optical rectification within, at the surface of, or beneath the gold nanostructures.

## Supplementary Note 2: Electromagnetic response

### 2.1 Preliminary investigation of lattice density and coupling effects

The resonance of plasmonic nanoantennas arranged in a regular array—with spacings comparable to the incident wavelength—is influenced by intermediate-/far-field coupling between resonators. This is known as the surface lattice resonance[4], which can be tuned to influence the resonance wavelength and Q factor compared with isolated nanostructures. Such effects are accounted for here in the electromagnetic simulations via periodic boundary conditions. Rather than iteratively optimize the resonator design, array geometry, and spacing for an 800 nm (or other desired) resonance, we simply optimize the resonator design in a 500 nm square lattice then test the effect of varying density (Fig. S3a). A shift in resonance position and quality factor is observed for the varying lattice constants, in good agreement with simulations (Fig. S3b). The 500 nm pitch metasurface exhibits the highest tip intensity enhancement (Fig. S3c) and THz emission (Fig. S3d) and is thus utilized for the present studies of the uniformly-oriented arrays. However, we note that significant further optimization is likely to be possible with different resonator designs, lattices, and nanoantenna spacings.

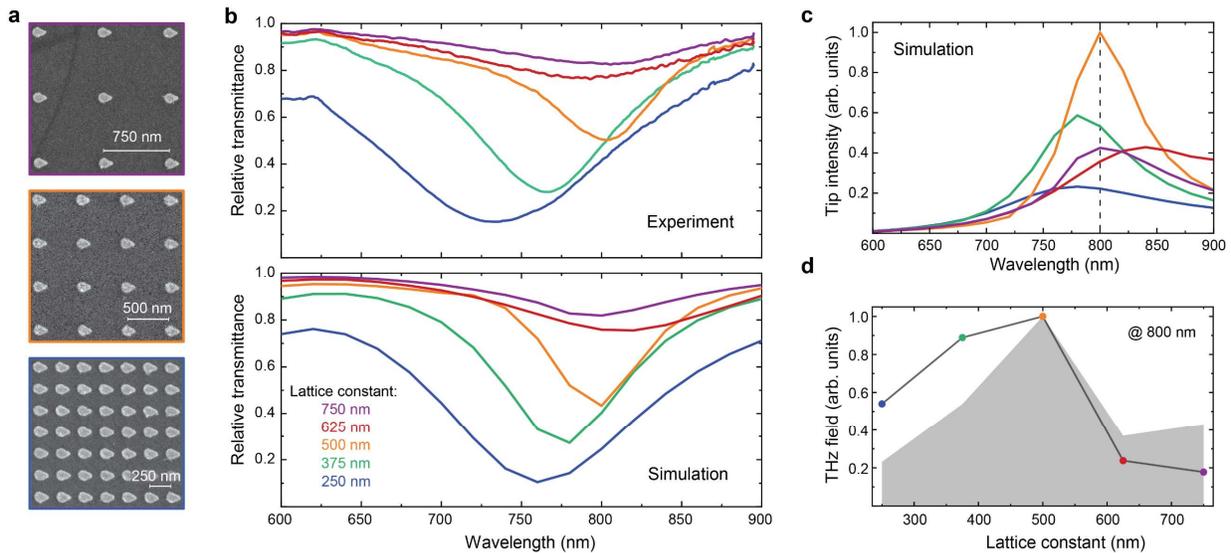

**Fig. S3 | Effect of metasurface density. a**, Electron micrographs of metasurfaces with 250 nm (bottom) to 750 nm (top) square lattice constant. **b**, Experimental (top) and simulated (bottom) relative transmittance spectra, correcting using the bare graphene/substrate as a reference. **c**, Simulated tip hot spot field intensity. **d**, Experimental (solid-dot line) THz field amplitude as a function of lattice constant, measured at 800 nm incident wavelength, compared with the simulated value from the density-scaled tip intensity (solid fill).

Because the same 800 nm nanoantenna design is utilized for the Kagome lattice, the surface lattice resonance is not optimized in these systems, leading to the lower field enhancements observed in Fig. 2 of the main text. Nevertheless, the linear response and properties of interest here are not affected. It is important to note that such lattice resonance effects will modify the plasmonic field enhancements but have little effect on the photocurrent distributions beyond the magnitude. In particular, unlike strong near-field coupling effects that can complicate the picture of linear superpositions of individual resonators, this intermediate/far-field coupling between elements of the sub-lattices do not appreciably modify the metasurface linear responses.

## 2.2 Graphene layer dependence

The effect of graphene layer number is tested by comparing the THz emission from metasurfaces fabricated on monolayer, bilayer, 3–5 layer, and 6–8 layer graphene (ACS Material), with SEM micrographs shown in Fig. S4a. An overall decreasing trend of THz emission with layer number is observed (Fig. S4b,c). This can be at least partially explained by increased damping with increase layer number, which leads to less absorbed power per layer (Fig. S4d) and thus less electronic heating, although the total absorbed power across all layers remains similar (Fig. S4e). Furthermore, the effect of metal doping will decrease away from the interfacial layer, leading to less acceleration of the hot carriers excited within deeper layers. Thus, the overall decrease in the current and THz emission signal is expected (simulation result in (Fig. S4c). Nevertheless, further work is needed to understand the various relevant contributions to the layer dependence, as well as ruling out systematics or sample-to-sample variations in the fabrication process for different numbers of layers. The different dispersion around the Dirac point for monolayer versus multilayer graphene are expected to have a small effect here, given the highly environmentally p-doped graphene (chemical potential far from the Dirac point, well into the Fermi liquid regime) and high photon energies (1.55 eV here). For the present studies, therefore, monolayer graphene appears to be the optimal choice.

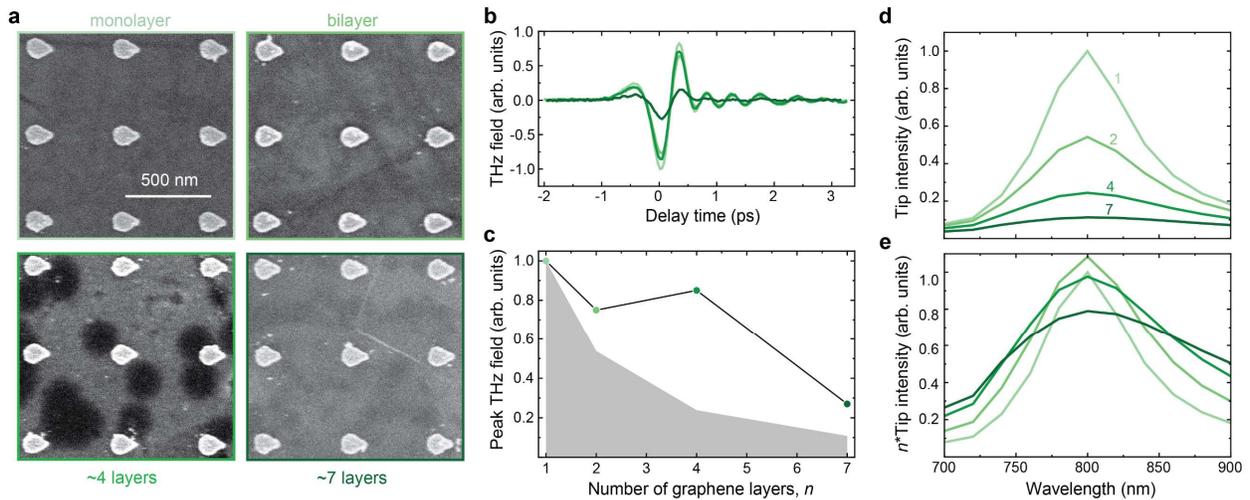

**Fig. S4 | Effect of graphene thickness. a**, Electron micrographs of metasurfaces fabricated on monolayer, bilayer, 3-5 (~4) layer, and 6-8 (~7) layer graphene. **b**, THz time domain traces for different layer numbers. **c**, Summary of the measured THz field amplitude (dash-dot line) compared with the peak simulated tip field intensity (solid fill) from panel d. **d**, Simulated field intensities at the nanoantenna tip, a few nanometers above the graphene interface. With graphene defined as a boundary layer, the layer number ($n$) is accounted for via modification of $\tilde{\sigma}_R \rightarrow n\tilde{\sigma}_R$. **e**, Same as panel d, but scaled by $n$.

## Supplementary Note 3: Thermodynamics

### 3.1 Two-temperature modeling

The full spatiotemporal evolution of electron and lattice temperatures in the metasurface unit cell requires a numerical treatment, as described in the main text (see Fig. 4, Methods, and Extended Data Fig. 5). However, the temperature evolution within bare graphene and individual nanostructures can be treated via simplified two-temperature modeling, neglecting the spatial degrees of freedom, so long as the absorption cross-sections are known. This offers additional insight on the role of the nanolocalized hot spot heating compared with uniform heating due to the 2.3% absorption in bare monolayer graphene. It also offers insight into the internal metal nanostructure heating, and whether these energy sources ought to also be accounted for in the hybrid system. Thus, here we implement the following simplified two-temperature kinetic model for the electronic ($T_e$) and phonon ($T_p$) temperatures, evaluated separately for the graphene and gold systems:

$$c_e(T_e)\frac{\partial T_e}{\partial t} = q_{\text{abs}}(t) - g_{\text{er}}(T_e - T_p), \tag{S1a}$$

$$c_p\frac{\partial T_p}{\partial t} = g_{\text{er}}(T_e - T_p). \tag{S1b}$$

Coupling with the substrate is neglected here, and it is approximated that the absorbed power from the optical pulse ($q_{\text{abs}}(t)$) is instantaneously transferred to the thermalized electron gas, bypassing the tens-of-femtosecond electron-electron thermalization times in both systems.

For graphene, we focus only on the energy transfer to the strongly-coupled optical phonons[5], with the electronic specific heat $c_e(T_e)$ described below (note 3.3), absorbed power $q_{\text{abs}}(t) = \frac{1}{2}\sigma_r E(t)^2$ (Methods), energy relaxation coupling constant $g_{\text{er}} = 2 \times 10^7$ WK$^{-1}$m$^{-2}$ (see note 3.2), and optical phonon specific heat $c_p$ taken from previous work[5]. The results are shown in Fig. S5a for 0.5 µJcm$^{-2}$ incident fluence and 100 fs pulse duration. Upon comparing the peak $T_e \approx 360$ K with the few-thousand Kelvin peak $T_e$ increase in the hybrid metasurface system (Fig. S8) under the same incident fluence, the role of the concentrated and enhanced power within the plasmonic hot spot in driving the photothermoelectric dynamics is further underscored. Note that at very short times before the excited carrier distribution has thermalized, $c_e$ is not strictly defined, but we maintain the instantaneous carrier–carrier thermalization approximation here. Meanwhile, in both the bare graphene and metasurface systems, the lattice (here meaning strongly-coupled optical phonon) temperature rise is minimal at only a few tens of Kelvin. See note 3.4 for calculations of the time-evolving $T_p$ within the metasurface unit cell.

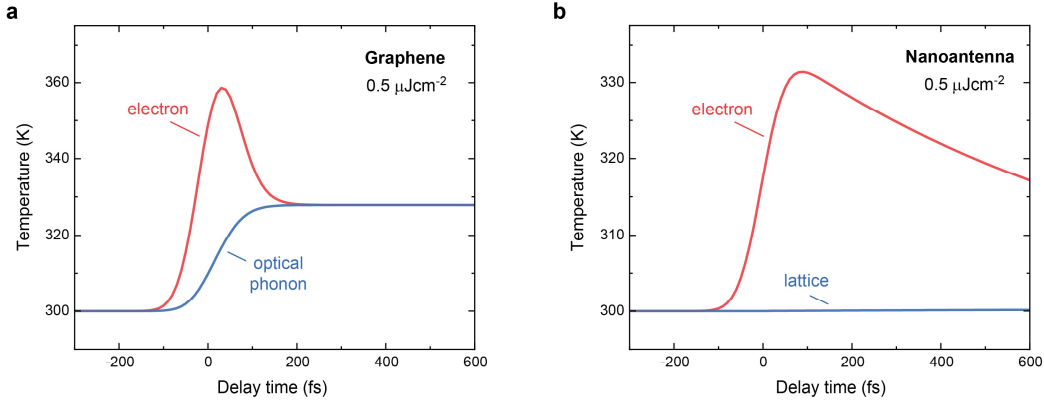

**Fig. S5 | Two-temperature modeling for bare graphene and gold nanoantennas. a**, Electron and optical phonon temperature evolution for bare graphene (no metasurface) under 100 fs pulse excitation centered around 0 delay time, neglecting acoustic phonon coupling. **b**, Electron and lattice (optical and acoustic phonon) temperature evolution for gold nanoantenna.

Now considering only the internal gold nanostructure thermal evolution, the relevant parameters are $c_e = \frac{\pi^2 k_B^2 n_e}{2\varepsilon_F} T_e$ (from Sommerfeld theory; $\varepsilon_F = 5.53$ eV and $n_e = 5.9 \times 10^{22}$ cm$^{-3}$), $q_{abs}(t) = \frac{1}{2Z_0 V}\alpha_{abs} E(t)^2$ (with $Z_0 = 377$ Ω the impedance of free space, $V = 3 \times 10^5$ nm$^3$ the nanoantenna volume, and $\alpha_{abs} = 4 \times 10^4$ nm$^2$ the simulated nanoantenna absorption cross section for resonant 800 nm excitation), $g_{er} = 2.5 \times 10^{16}$ W K$^{-1}$m$^{-3}$ is the electron-phonon coupling constant[6], and $c_l = 2.4 \times 10^6$ J K$^{-1}$m$^{-3}$ for gold. The results are shown in Fig. S5b for 0.5 μJ cm$^{-2}$ incident fluence. For this system, the electronic temperature increase is an even more modest ~30 K and the lattice temperature increase is negligible. Thus, the gold may serve as a thermal sink (unaccounted for in the present work) for the local hot carrier dynamics within the graphene around the tip hot spot but would not be sufficiently hot to contribute as a source in any of the dynamics.

### 3.2 Ultrafast dynamics measured via transient reflectivity

We perform transient reflectivity measurements to characterize the coupling constant ($g_{er}$) between the electronic and lattice systems in our graphene samples. For these measurements we utilize collinear, cross-polarized 800 nm + 800 nm pump-probe microscopy to measure transient (sub-picosecond) changes in the graphene reflectivity. Pump and probe pulses from a Coherent Vitesse oscillator (80 MHz repetition rate) are focused through a Mitutoyo 20× objective onto the sample under high vacuum (~10$^{-6}$ Torr; Janis ST500 optical cryostat). The reflected probe beam is collected on a Si photodiode, with pump filtered out via a linear polarizer. A lock-in amplifier (SR830) is utilized to read out the modulated probe signal induced by the 2.5 kHz-chopped pump beam. The transient reflectivity traces (Fig. S6a) reveal two well-known general behaviors of graphene[7,8]: a fast (~400 fs) decay due to electron–optical phonon scattering and a slower (~1.1 ps) decay due to thermalization with acoustic phonons (acoustic phonon bottleneck). The slower thermalization with the acoustic phonon system can proceed via optical–acoustic phonon–phonon coupling or directly via disorder-assisted electron–acoustic phonon scattering[9].

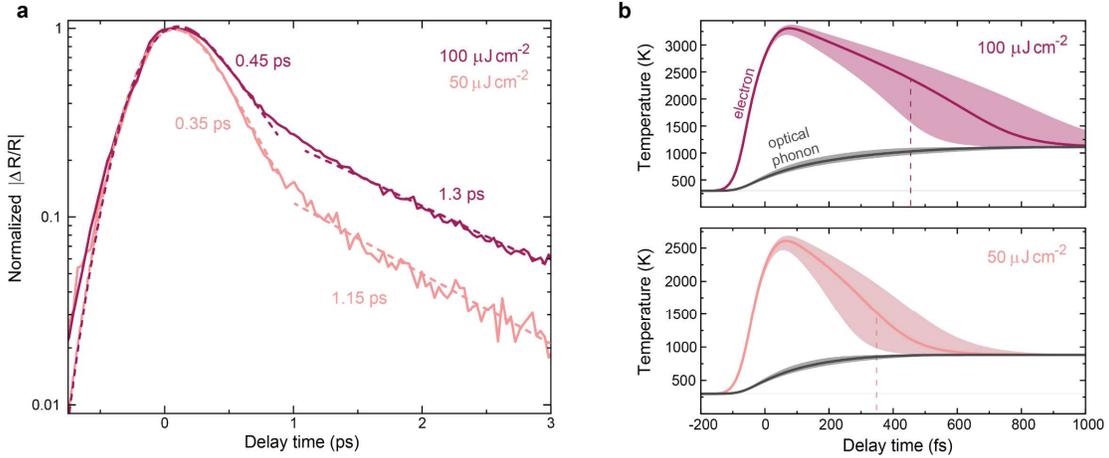

**Fig. S6 | Ultrafast transient reflectivity of monolayer graphene and electron-optical phonon coupling constant. a**, Transient reflectivity time traces for 50 µJcm⁻² and 100 µJcm⁻² incident fluence (solid lines). Dashed lines indicate fits to electron-optical phonon thermalization timescales convolved with instrument response function, with separate exponential fits to acoustic phonon thermalization timescales. **b**, Two-temperature modeling for $g_{er} = 2 \times 10^7$ WK⁻¹m⁻², with 30% uncertainty bounds from $g_{er} = 1.3 \times 10^7$ to $2.7 \times 10^7$ shown (solid fill). Dashed lines mark the measured decay times.

For suitably high signal-to-noise measurements on the monolayer graphene (with peak $\frac{\Delta R}{R} \sim 0.5 \times 10^{-3}$), we work at high fluences up to 100 µJ cm⁻². These values also approach the locally plasmon-enhanced fluences in the metasurface system, but with stronger overall heating without the diffusive cooling pathway. The coupling parameter, $g_{er}$, is tuned for best agreement between the measured fluence-dependent electron–optical phonon thermalization times and two-temperature models (Fig. S6b). We estimate $g_{er} \approx 2 \times 10^7$ WK⁻¹m⁻² based on these model assumptions and the high-temperature behavior of graphene discussed in the next section.

## 3.3 High-temperature properties of graphene

Under ultrafast exposure, the electronic system is transiently far from equilibrium with the lattice, reaching temperatures of several thousand Kelvin while the lattice remains near room temperature (thus avoiding issues such as electrode melting that would occur at high temperatures in studies around equilibrium). Expressions commonly utilized at low temperatures or up around room temperature will no longer be valid as $T \to T_F$ ($T_F = 3600$ K for $\varepsilon_F = -300$ meV). We thus include a discussion on the behavior of several important quantities in the high-temperature regime.

The $T_e$-dependent electronic specific heat is critical for determining the heating of the electronic system, and can be written generically as

$$c_e(T_e) = \int_{-\infty}^{\infty} d\varepsilon \, \varepsilon \rho(\varepsilon) \frac{\partial f(\varepsilon, T_e)}{\partial T_e}, \tag{S3}$$

where $\rho(\varepsilon) = \frac{2|\varepsilon|}{\pi \hbar^2 u_F^2}$ is the electronic density of states and $f(\varepsilon, T_e) = \left(e^{(\varepsilon - \mu(T_e))/k_B T_e} + 1\right)^{-1}$ is the Fermi-Dirac distribution. When the temperature dependence of the chemical potential is accounted for, this $c_e$ interpolates between the Fermi liquid ($k_B T_e \ll |\mu|$) and Dirac fluid ($k_B T_e \gg |\mu|$) expressions for the specific heat[10],

$$c_e^{\text{FL}}(T_e) = \frac{2\pi k_B^2 \varepsilon_F}{3 \hbar^2 u_F^2} T_e, \tag{S4a}$$

$$c_e^{\text{DF}}(T_e) = \frac{18 \zeta(3) k_B^3}{\pi \hbar^2 u_F^2} T_e^2, \tag{S4b}$$

where $\zeta(3) = 1.202$ is the Reimann-Zeta function. The specific heats in each regime are plotted in Fig. S7a by comparison with the result from Eq. S3. The $T_e$-dependent chemical potential—which necessarily varies to conserve charge—is given in both regimes by[10,11],

$$\mu^{\text{FL}}(T_e) = \varepsilon_F \left(1 - \frac{\pi^2 k_B^2 T_e^2}{6 \varepsilon_F^2}\right), \tag{S5a}$$

$$\mu^{\text{DF}}(T_e) = \frac{\varepsilon_F^2}{4 \ln(2) k_B T_e}, \tag{S5b}$$

and here we simply interpolate between these two limits (Fig. S7a) as

$$\mu(T_e) \approx \left(1 + \frac{T_e}{T_F/2}\right)^{-6} \mu^{\text{FL}} + \left(1 - \left(1 + \frac{T_e}{T_F/2}\right)^{-6}\right) \mu^{\text{DF}}, \tag{S6}$$

positioned around $T_F/2$ (half the Fermi temperature, $T_F = \varepsilon_F/k_B$) to preclude the sign change that occurs in Eq. S5a as $T_e \to T_F$.

The electron-electron scattering rate ($\tau_{ee}^{-1}$) in graphene in the Fermi liquid regime scales as $-T_e^2 \ln(\frac{k_B T_e}{|\varepsilon_F|})$ at low temperatures[12] ($T_e \ll T_F$), but this expression becomes problematic as $T_e \to T_F$ ($\tau_{ee}^{-1} \to 0$ and then becomes negative). The high-temperature behavior can be evaluated using Fermi's golden rule[13],

$$\tau_{ee}^{-1}(\varepsilon_i, T_e) = \frac{2\pi}{\hbar} M^2 \int_{-\infty}^{\varepsilon_i} d\varepsilon_f \left(1 - f(\varepsilon_f)\right) \rho(\varepsilon_f)$$

$$\times \int_{-\infty}^{\infty} d\varepsilon_i' \, f(\varepsilon_i') \rho(\varepsilon_i') \left(1 - f(\varepsilon_i' + \Delta)\right) \rho(\varepsilon_i' + \Delta), \tag{S7}$$

where $\Delta = \varepsilon_i - \varepsilon_f$, $M$ is the transition matrix element (approximated as constant), subscripts indicate initial ($i$) and final ($f$) states, and the primed (unprimed) energies refer to the secondary (primary) electrons. The approximation of constant $M$ neglects the onset of interband transitions at high temperatures, but this simplifying assumption is suitable for the purposes of the present discussion. Integration of Eq. S7 yields a nearly quadratic energy dependence for the Fermi liquid, $\tau_{ee}^{-1} \sim (\varepsilon - \varepsilon_F)^2$, modified at low energies ($|\varepsilon_i - \varepsilon_F| < k_B T_e$) due to the finite temperature. The overall $T_e$ dependence is then determined by averaging over energy,

$$\tau_{ee}^{-1}(T_e) = \int_{-\infty}^{\infty} d\varepsilon_i \, \tau_{ee}^{-1}(\varepsilon_i, T_e) \frac{\partial f}{\partial \varepsilon_i} \,. \tag{S8}$$

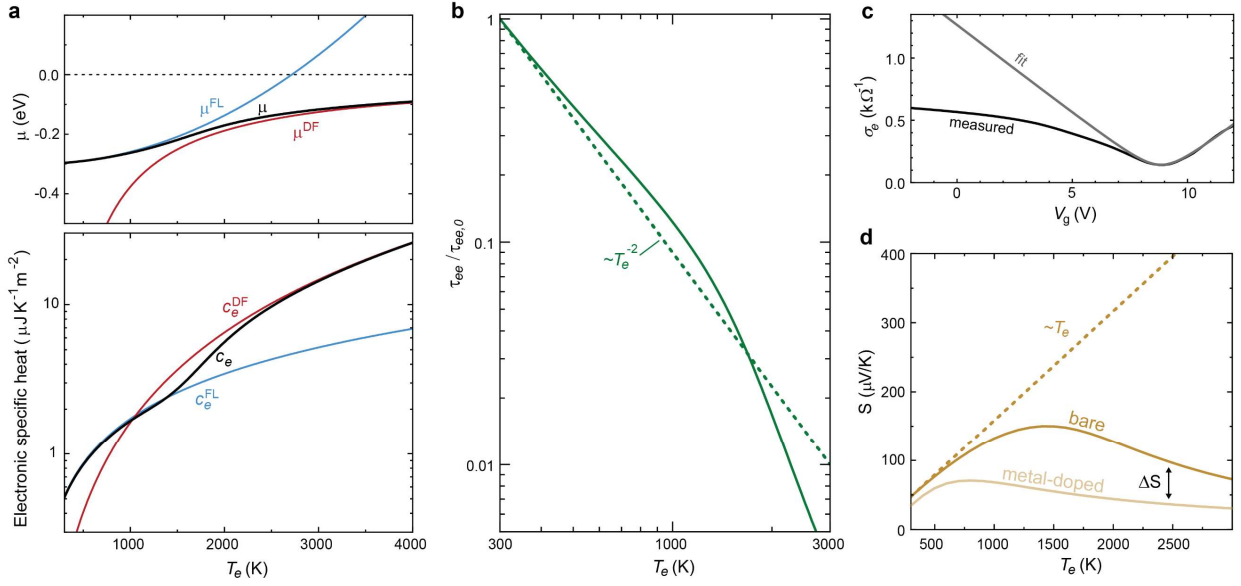

**Fig. S7 | High-$T_e$ calculations for graphene. a**, Chemical potential (top) and specific heat (bottom). **b**, Scattering time relative to $T_0$ value compared with standard $T_e^{-2}$ Fermi liquid theory behavior. **c**, Measured conductivity for graphene device and idealized fit excluding extrinsic effects, utilized for calculations of the Seebeck coefficient. **d**, Seebeck coefficient calculated for the bare and metal-doped graphene regions based on Eq. S9, compared with the linear-$T_e$ behavior that prevails at low temperatures.

The result is plotted in Fig. S7b, revealing $\tau_{ee}^{-1}(T_e) \sim T_e^2$ dependence. This behavior is responsible for locally driving the electronic system around the nanoantennas into an apparently hydrodynamic regime, where $\tau_{ee} \ll \tau_{mr}$, influencing the spatially-varying viscosity, $\nu \sim \tau_{ee}$. The room temperature value of $\tau_{ee}$ was determined to be $\sim$200 fs in previous work[14].

The Seebeck coefficient at high temperatures can be determined from the generalized form of the Mott relation[15,16],

$$S(\mu) = -\frac{1}{eT_e} \frac{\int_{-\infty}^{\infty} d\varepsilon \, (\varepsilon - \mu) \sigma_e(\varepsilon) \frac{\partial f}{\partial \varepsilon}}{\int_{-\infty}^{\infty} d\varepsilon \, \sigma(\varepsilon) \frac{\partial f}{\partial \varepsilon}}, \tag{S9}$$

The measured electrical conductivity is fit to an idealized functional form[17,18], $\sigma_e(\mu) = \sigma_{\min}\left(1 + \frac{\mu^4}{w^4}\right)^{-2}$ where $\sigma_{\min}$ is the minimum conductivity and $w = 100$ meV is the width of the charge neutrality region (Fig. S7c). This removes extrinsic contributions that do not influence the local conductivity at the nanoantenna tips. We approximate $\sigma_e$ (which also appears in the Wiedemann-Franz expression for $\kappa_e$) to be independent of $T_e$, as Umklapp scattering is suppressed in graphene. At the high temperatures relevant to the present studies, Eq. S9 yields a sublinear and even nonmonotonic dependence on $T_e$, plotted in Fig. S7d for both the bare and gold-doped graphene regions. For the bare graphene, Eq. 6 (Fig. S7a) is utilized for $\mu_{\text{bare}}(T_e)$, while we approximate $\mu_{\text{pinned}} \approx -50$ meV as a constant pinned by the gold chemical potential. The resulting $\Delta S(T_e)$ is then responsible for the photothermoelectric acceleration of charge.

### 3.4 Thermodynamic energy flow

The energy cascade in graphene following ultrafast optical excitation generally involves as (i) rapid carrier-carrier thermalization within ~100 fs, (ii) thermalization with optical phonons within a few hundred femtoseconds, and (iii) full thermalization with acoustic phonons on $> 1$ ps timescales[7,8,10,11,19,20]. As mentioned above, thermalization with the acoustic phonon system can be accelerated via a disorder-assisted (so-called supercollision) cooling channel, directly between the electrons and acoustic phonons[9,10]. If photon energy $> 2\varepsilon_F$ such that interband transitions are allowed (as is the case here), separated Fermi-Dirac distributions for the electron and hole gases can be established on tens-of-femtosecond timescales before a single electronic Fermi-Dirac distribution is established on < 150 fs timescales[11,19]. In the modeling performed here, we approximate an instantaneous transfer of energy between the optical pulse and the fully thermalized electronic distribution. While the convolution of these ~100 fs dynamics with the ~100 fs optical pulse duration can be expected to skew the short-timescale behaviors, such effects will have little bearing on the overall thermodynamic energy and hydrodynamic momentum flows examined here.

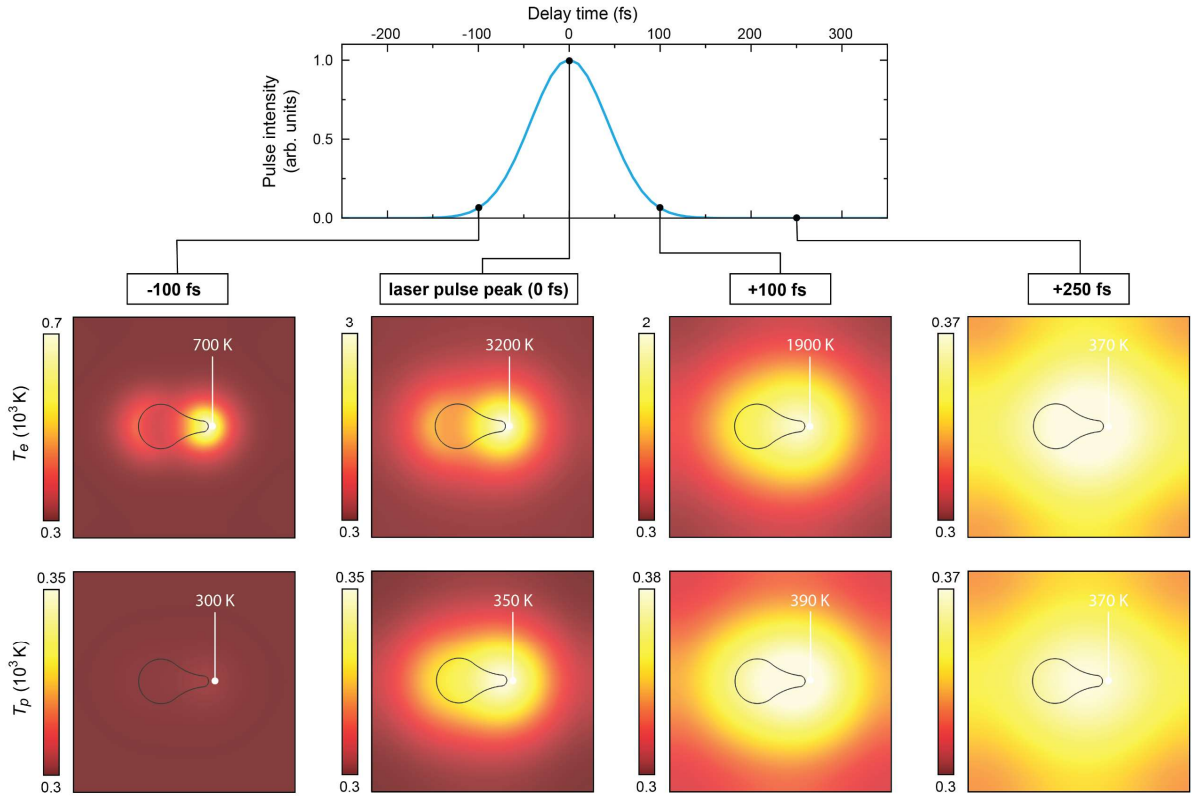

**Fig. S8 | Electronic and optical phonon temperature evolution in graphene.** Spatial temperature distributions at various times relative to the incident laser pulse, assuming instantaneous transfer of energy into a locally thermalized hot carrier distribution.

In some previous works[5,19], the strong coupling between the electronic and optical phonon systems has led to rapid optical phonon heating up to > 1000 K. These works, however, employ 30–300-fold higher fluences than typically utilized in the present studies (< 1 µJ cm$^{-2}$) on bare graphene. The total absorbed power here is modest, with electronic superheating only occurring due to the highly concentrated plasmonic field distribution. As a result of this relatively low overall heat load, fast electronic thermal diffusion, and the delayed energy transfer to the optical phonon system, the peak optical phonon temperature within the unit cell is calculated to be merely 390 K (Fig. S8). Coupling to the acoustic phonon system on longer timescales is neglected here. The strongly heated electronic system (with dramatically faster $\tau_{ee}$) and near-room-temperature phonon system (with little change to $\tau_{mr}$) thus supports the view of a transient light-induced hydrodynamic phase. The resulting charge flows are discussed in the main text and examined further in the next section.

## Supplementary Note 4: Hydrodynamics

### 4.1 Quantitative comparison with experiments

Quantitative results for simulated DC photocurrents (averaging the ultrafast time-evolving current flow) are shown as a function of incident laser fluence in Fig. S9. Comparison with experimental values shows good agreement, to within a factor of 2. Note that the applicability of the hydrodynamic description will break down at low incident fluences (low temperatures) due to insufficient electronic heating. The major sources of uncertainty in the model include the following: (i) The precise plasmonic field enhancement within the graphene layer, due to challenges in classical (and idealized) simulations of interfaces. (ii) The spatial extent of the transition region between the metal-doped and bare graphene regions, which is taken here to be

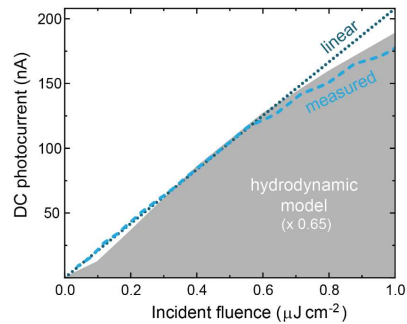

**Fig. S9 | Incident fluence dependence.** Measured fluence-dependent DC photocurrent (dashed blue line) versus values calculated via hydrodynamic model (solid gray fill). The dotted line shows a best-fit linear dependence to the experimental data at low fluence.

~10 nm via Gaussian convolution over the two domains. (iii) The effects of nonuniform charge density. (iv) High-temperature behaviors in the thermodynamic and hydrodynamic parameters. Despite these sources of uncertainty, the simulated flow behaviors remain robust to changes in the exact functional forms and values of the heat capacities, thermal conductivities, thermopower, etc. Therefore, the good agreement between experiment and modeling—based on the best approximations described above and using no ad hoc parameter variation/optimization—suggests that such simulations are a reasonable starting point for understanding and predicting the nanoscale flow behaviors occurring in our system.

Calculated polarization dependence is also in good agreement with experiments (and the analysis based on a simple linear response), as shown in Fig. S10. Compared with the simple linear responses and superposition of currents from the different nanoantennas described in the main text, the hydrodynamic simulations also account for coupling that occurs due to the incompressibility constraint. Even so, the symmetry of the light-matter interaction evidently prevails.

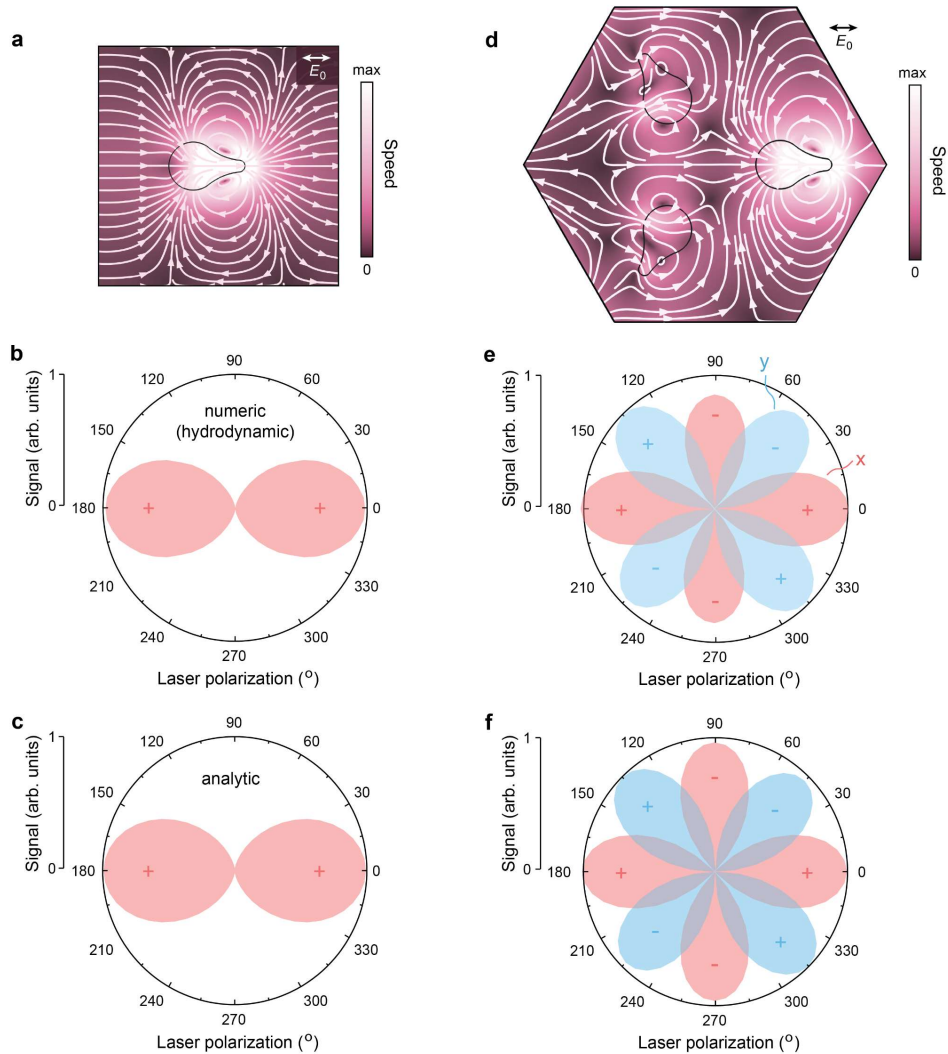

**Fig. S10 | Comparison between analytic and hydrodynamic polarization dependence. a**, Flow profile for the unit cell f the square lattice. **b**, Polarization dependence of the photocurrent (*x* component; *y* component is negligibly small) calculated via hydrodynamic model. **c**, Polarization dependence of the photocurrent calculated analytically for a linear response (*x* component; *y* component is identically zero by symmetry). **d–f**, Same as panels a–c but for the Kagome lattice unit cell (now including *y* photocurrent component).

**4.2 Chirality and magnetization**

Most of the metasurfaces studied here have been achiral, with the exception of the perturbed Kagome lattice (Extended Data Fig. 2) with local planar chirality, and the azimuthal metasurface (Fig. 3) with global planar chirality. Transient spatially-varying magnetic fields will accompany all of the current flow profiles, though the introduction of structural chirality (i.e., with no mirror symmetry planes perpendicular to the metasurface) will introduce chiral orbital nanocurrents and net orbital magnetization. A simple example of this is shown in Fig. S11. Such systems may prove useful for controlling magnetic interactions within various symmetry-broken metasurfaces or in nearby materials, although a simple estimate limits the transient magnetic fields to the µT or low-mT ranges in graphene. This limit is imposed by the current density, $\mathbf{j} = e n_e \mathbf{u}$, where $u < u_F$ (and $\ll u_F$ under most experimentally accessible conditions) and $n_e \lesssim 10^{13}$ cm$^{-2}$ by electrostatic doping.

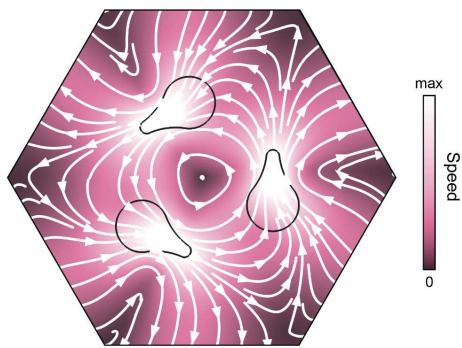

**Fig. S11 | Local vortical flow and net magnetization in a chiral unit cell.** An example of a chiral nanostructure arrangement with corresponding counter-clockwise hydrodynamic flow, serving as a local magnetic dipole.

# Supplementary References

1   Li, X. S. *et al.* Large-area synthesis of high-quality and uniform graphene films on copper foils. *Science* **324**, 1312-1314 (2009).

2   Wang, H. M., Wu, Y. H., Cong, C. X., Shang, J. Z. & Yu, T. Hysteresis of electronic transport in graphene transistors. *ACS Nano* **4**, 7221-7228 (2010).

3   Novoselov, K. S. *et al.* Two-dimensional gas of massless Dirac fermions in graphene. *Nature* **438**, 197-200 (2005).

4   Kravets, V. G., Kabashin, A. V., Barnes, W. L. & Grigorenko, A. N. Plasmonic surface lattice resonances: A review of properties and applications. *Chem. Rev.* **118**, 5912-5951 (2018).

5   Lui, C. H., Mak, K. F., Shan, J. & Heinz, T. F. Ultrafast photoluminescence from graphene. *Phys. Rev. Lett.* **105**, 127404 (2010).

6   Groeneveld, R. H. M., Sprik, R. & Lagendijk, A. Femtosecond spectroscopy of electron-electron and electron-phonon energy relaxation in Ag and Au. *Phys. Rev. B* **51**, 11433-11445 (1995).

7   Breusing, M. *et al.* Ultrafast nonequilibrium carrier dynamics in a single graphene layer. *Phys. Rev. B* **83**, 153410 (2011).

8   Wang, H. N. *et al.* Ultrafast relaxation dynamics of hot optical phonons in graphene. *Appl. Phys. Lett.* **96**, 081917 (2010).

9   Song, J. C. W., Reizer, M. Y. & Levitov, L. S. Disorder-assisted electron-phonon scattering and cooling pathways in graphene. *Phys. Rev. Lett.* **109**, 106602 (2012).

10  Massicotte, M., Soavi, G., Principi, A. & Tielrooij, K. J. Hot carriers in graphene - fundamentals and applications. *Nanoscale* **13**, 8376-8411 (2021).

11  Gierz, I. *et al.* Snapshots of non-equilibrium Dirac carrier distributions in graphene. *Nat. Mater.* **12**, 1119-1124 (2013).

12  Giuliani, G. F. & Quinn, J. J. Lifetime of a quasiparticle in a two-dimensional electron-gas. *Phys. Rev. B* **26**, 4421-4429 (1982).

13  Zarate, E., Apell, P. & Echenique, P. M. Calculation of low-energy-electron lifetimes. *Phys. Rev. B* **60**, 2326-2332 (1999).

14  Bandurin, D. A. *et al.* Negative local resistance caused by viscous electron backflow in graphene. *Science* **351**, 1055-1058 (2016).

15  Ashcroft, N. W. & Mermin, N. D. *Solid state physics*. (Saunders College, 1976).

16  Ghahari, F. *et al.* Enhanced thermoelectric power in graphene: Violation of the Mott relation by inelastic scattering. *Phys. Rev. Lett.* **116**, 136802 (2016).

17  Shiue, R. J. *et al.* High-responsivity graphene-boron nitride photodetector and autocorrelator in a silicon photonic integrated circuit. *Nano Lett.* **15**, 7288-7293 (2015).

18  Shautsova, V. *et al.* Plasmon induced thermoelectric effect in graphene. *Nat. Commun.* **9**, 5190 (2018).

19  Johannsen, J. C. *et al.* Direct view of hot carrier dynamics in graphene. *Phys. Rev. Lett.* **111**, 027403 (2013).

20  Brida, D. *et al.* Ultrafast collinear scattering and carrier multiplication in graphene. *Nat. Commun.* **4**, 1987 (2013).